\documentclass[12pt]{article}
\usepackage{amssymb,amsmath,latexsym,graphicx, bm, mathrsfs, hyperref, enumitem}

\begin{document}

\title{Dynamical systems of null geodesics \\ and solutions of Tomimatsu-Sato 2 \vspace{-0.5cm} \author{ Sumanto Chanda, Partha Guha$$ \\ \textit{S.N. Bose National Centre for Basic Sciences}\\ \textit{JD Block, Sector-3, Salt Lake, Calcutta-700098, INDIA.} \vspace{-0.4cm} \date{\today}}}

\maketitle
\thispagestyle{empty}

\vspace{-1.25cm}

\begin{center}
\texttt{\small sumanto12@bose.res.in ,  partha@bose.res.in}
\end{center}

\vspace{-0.6cm}
\begin{center}
{\it Dedicated with admiration, gratitude and deep respect \\ to Professor Gary Gibbons on his 
70th birthday}
\end{center}

\vspace{-0.25cm}


\smallskip

{\bf{Keywords and keyphrases :}} Null-geodesic, Optical-mechanical formulation, Binet equation, 

Kerr metric, Tomimatsu-Sato metric, Kepler-Hooke duality, Chebyshev theorem. 

\bigskip

\vspace{-0.5cm}
\abstract{We have studied optical metrics via null geodesics and optical-mechanical formulation of classical 
mechanics, and described the geometry and optics of mechanical systems with drag dependent quadratically 
on velocity. Then we studied null geodesics as a central force system, deduced the related Binet's 
equation applied the analysis to other solutions of Einstein's equations in spherically symmetric spaces, paying 
special attention to the Tomimatsu-Sato metric. Finally, we examined the dualities between different 
systems arising from conformal transformations that preserve the Jacobi metric.}

\vspace{-0.5cm}
\tableofcontents

\newpage

\setcounter{page}{1}
\numberwithin{equation}{section}

\setlength{\topmargin}{-0.6in}
\setlength{\textheight}{9.2in}

\section{Introduction}

Optical metrics are essentially null geodesics in a given spacetime. Such null arcs can be studied 
by projecting the curves onto lower dimensional spatial surfaces. As example, if a 
metric admits a timelike Killing vector $K$ orthogonal to a hypersurface, the null geodesic will 
project down to unparameterised geodesics of the optical metric on the space of orbits of 
$K$. Similar constructions were studied for metrics admitting a stationary Killing vector \cite{ghww} 
or a timelike conformal retraction \cite{casey2} where the projected null curves provide some notion of 
geometric structure to a hypersurface. Because the null metric vanishes, the geodesics are defined by 
minimising only the spatial part of the metric.

Geodesics on null curves are formulated in accordance with Fermat's principle, as the spacetime 
curves cannot be minimized. One direct utility is in observational astronomy in the 
study of gravitational lensing. Using null geodesics, one can interpret gravitational fields as 
transparent media with a refractive index. Conversely, one can also view transparent media as 
regions with localised gravitational fields that exist around each and every individual molecule, 
as speculated by P. de Fermat and P.L. Maupertuis \cite{maup}. 

Since we are discussing null-geodesic formulation using only the spatial part of the metric, 
we must also consider the Jacobi metric \cite{gwg, cgg,cgg1}, which is a reduction of the stationary 
geodesic metric to just the spatial part. The analytical calculations involving null geodesics 
in spherically symmetric spacetimes using Weierstrass elliptic functions are given in \cite{gwg1}. Preservation 
of the classical Jacobi metric under conformal transformation can lead us to interesting dual 
pairs of mechanical systems, such as the Bohlin transformation \cite{dual, mls}, as shown by Casey in 
\cite{casey1}. These dual systems can be written in the form of Binet's equations, casting them as 
central force systems for particular potentials, allowing us to apply dynamical solutions to 
study black hole optics. One way to provide solutions to Binet's equation is to solve the null 
geodesic equation using Chebyshev's theorem \cite{cgly, mz, plmath}. We also briefly mention the anisotropic 
case, the optical anisotropy of curved space is demonstrated by means of a rigorous algebraic 
analysis in \cite{KR}. 

Most practical mechanical problems deal with drag that occur during their operation. 
Performing an optical mechanical formulation for such systems helps generalize our analysis. 
The metric related to such mechanical systems describe a more general category of spacetimes.

The Tomimatsu-Sato metric discovered in 1972 by Tomimatsu and Sato \cite{tomsato, tomsatop} describes 
solutions for stationary axisymmetric system. It has a ``naked singularity" outside the non-
regular Killing horizon which distinguishes it from the Kerr metric. Bose and Wang in \cite{bosewang} 
have provided solutions for the null geodesics under approximations, while various limits of 
the metric were discussed by Kinnersley and Kelley in \cite{kinkel}. It would be interesting to extend 
the solutions of the null geodesics to these limits. \bigskip

In this article, we have explored the optical nature of spacetime via null geodesics and 
their interpretation as dynamical systems. We demonstrated how optics derive from null 
geodesics and applied the theory to one example. Then we deduced the spacetime metric from 
mechanical system for two types of velocity-dependent drag and studied their optics. Casey's 
analysis of null geodesics as a central force system for Schwarzschild-Tangherlini metric and 
its trajectory solutions via Binet's equation was extended to general solutions of Einstein's 
equations, and applied to two other spacetimes. Finally,  we studied duality of such solutions 
under conformal transformation, such as Bohlin-Arnold duality, implying conformal duality of 
their null-geodesics. The article has been organized as follows. 

We shall first study the preliminaries on null geodesics, derive Snell's Law used in refractive 
optics, and deduce the refractive indices of the Kerr metric. Then, we shall write the relativistic 
and non-relativistic Jacobi metric in optical-mechanical form. Furthermore,we study how 
mechanics of more general isotropic spaces involve drag, and deduce the spacetime metric and 
optics from a damped equation of motion. 

In the next section, we shall extrapolate S. Casey's results for $n$-dimensional Schwarzschild 
metrics \cite{casey1} to general spherically symmetric metric. We shall first formulate the metric for 
general spherically symmetric spaces. Then we shall describe the null geodesics as a two 
dimensional central force system, and deduce the related Binet's equations. Then we proceed to 
deduce solutions to the Binet's equations from the null geodesic equations under approximations 
and binomial expansions in terms of beta functions, both complete and incomplete. This 
analysis of null-geodesics is then applied to other examples of solutions to Einstein's equations 
and Kerr spacetimes, and special emphasis paid to the Tomimatsu-Sato metric. Null geodesic 
solutions to some of the limits of the Tomimatsu-Sato metric were deduced.

Following that, we shall explore the various mechanical dualities that arise from 
preservation of the classical Jacobi metric under a conformal co-ordinate map, one example being 
Bohlin's transformation.

\section{Preliminaries: Null-geodesics} \label{sec:sec2}

It was speculated P.L. Maupertuis in \cite{maup} how the refraction of light upon passing into a medium 
could be due to gravitational effects. Furthermore, null-geodesics are unique since the speed 
of a particle (photon) travelling along a null-geodesic remains unchanged under local Lorentz 
group transformations. In special relativity, for flat spaces, this leads to Einstein's postulate 
about the universality of the speed of light in all inertial frames, which holds true locally, even 
in refracting media. \bigskip

Here, we will demonstrate with a null geodesic in isotropic space how refractive phenomena 
can arise from a gravitational metric, as shown in \cite{br}. Suppose we have an isotropic space-time 
metric given by:
\begin{equation}
\label{isotrop} ds^2 = A ({\vec r}) c^2 dt^2 - B ({\vec r}) |d {\vec r}|^2 \qquad \Rightarrow \qquad \left( \frac{d s}{d \tau} \right)^2 = A ({\vec r}) c^2 {\dot t}^2 - B ({\vec r}) |\dot{\vec r}|^2.
\end{equation}
From the null geodesic equation of (\ref{isotrop}), the local refractive index $n(\vec r)$ wrt vaccum is:
\begin{equation}
\label{refindex} ds^2 = 0 \qquad \Rightarrow \qquad n ({\vec r}) = \frac c{v_{null}} = \sqrt{\frac{B ({\vec r})}{A ({\vec r})}}, \qquad \qquad \text{where } \ v = \frac{|\dot{\vec r}|}{\dot t}.
\end{equation}
where we can see that so long as the position is unchanged, the speed is universal in all inertial frames of reference. 
$$ds^2 = 0 \qquad \Rightarrow \qquad A ({\vec r}) c^2 dt^2 - B ({\vec r}) |d {\vec r}|^2 \ = A ({\vec r}) c^2 d \tau^2 - B ({\vec r}) |d {\vec \rho}|^2 \ = \ 0,$$
$$\Rightarrow \qquad \left| \frac{d \vec r}{d t} \right| = \left| \frac{d \vec \rho}{d \tau} \right| = c \sqrt{\frac{A ({\vec r})}{B ({\vec r})}}.$$
$\tau$ and $\vec \rho$ being time and position vector co-ordinates in another local Lorentz frame. The 
solutions to Einstein's equations for vaccum usually have $A B = 1$, for which the refractive 
index (\ref{refindex}) becomes $n ({\vec r}) = \left( A ({\vec r}) \right)^{-1}$. \smallskip

For anisotropic spaces, each individual spatial direction has its own refractive index given by $n_i ({\vec r})$. Here the stationary metric, where $g_{ij}$ is assumed to have been diagonalised via similarity transformation, is written as
\begin{equation}
\label{niso} ds^2 = h_{00} (x) c^2 dt^2 - 2 h_{i0} (x) c \ d t \ dx^i - h_{ij} (x) dx^i dx^j.
\end{equation}
where the metric components given above $h_{00}, h_{ij} > 0 \ \forall \ i, j$ are positive definite. The refractive index 
along each direction can be deduced by setting all other co-ordinates constant at a time:
\begin{equation}
\label{dipnull} ds^2 = 0, \quad x^j = const. \quad \forall \ j \neq i \qquad \Rightarrow \qquad h_{00} ({\vec r}) c^2 - 2 h_{i0} ({\vec r}) c \ v^i_{null} - h_{ii} ({\vec r}) \left( v^i_{null} \right)^2 = 0.
\end{equation}
 If we choose to write (\ref{dipnull}) as a quadratic equation for the refractive index $n_i$, we will have the following equation, with two solutions:
\begin{equation}
\label{solutions} h_{00} ({\vec r}) (n_i)^2 - 2 h_{i0} ({\vec r}) n_i - h_{ii} ({\vec r}) = 0 \qquad \Rightarrow \qquad n_i = \frac c{v^i_{null}} = \frac{h_{i0} \pm \sqrt{(h_{i0})^2 + h_{00} h_{ii}}}{h_{00}}.
\end{equation}
It is clearly evident that if $h_{ii}$ and $h_{00}$ are positive definite, then $\sqrt{(h_{i0})^2 + h_{00} h_{ii}} > |h_{i0}|$. Thus, 
regardless of the signature of $h_{i0}$ (ie. $h_{i0} = \pm |h_{i0}|$), choosing the $-$ sign option in (\ref{solutions}), gives 
a negative refractive index (ie. $n_i < 0 \ ;$ for $ h_{i0} = \pm |h_{i0}|$), technically implying that light is 
travelling in a direction opposite to the direction it would take in vaccum. To consider only the realistic solution, we shall 
take only the $+$ sign option in (\ref{solutions}), which means:
\begin{equation}
\label{nisoindex} n_i = \frac{h_{i0} + \sqrt{(h_{i0})^2 + h_{00} h_{ii}}}{h_{00}}.
\end{equation}
We must note that negative refractive indices are permissible when considering 
metamaterials \cite{fnb, cs, csl}.

Since we have constrained the length of a null geodesic to vanish, applying a variational 
process upon it's length seems futile. It is more sensible to vary the spatial part alone, 
effectively applying Fermat's principle of light travelling by the shortest path between two points. 
Thus, using (\ref{refindex}) we can say that the optical arc integral, and its Euler-Lagrange equation are
\begin{equation} \label{optlag} 
\begin{split}
l = \int_1^2 d \tau \ c {\dot t} &= \int_1^2 d \tau \ n ({\vec r}) |\dot{\vec r}| = \int_1^2 n ({\vec r}) |d \vec r| \qquad \qquad L_{null} = c \dot{t} = n ({\vec r}) |\dot{\vec r}| , \\
\delta l &= 0 \qquad \Rightarrow \qquad \frac{d \ }{d \tau} \left( n ({\vec r}) \frac{\dot{\vec r}}{|\dot{\vec r}|} \right) = |\dot{\vec r}| {\vec \nabla} n ({\vec r}).
\end{split}
\end{equation}
We will regard the arc length as a natural parameter along the curve. If we parametrize with respect to arc length, the reparametrized velocity can be written as a 
unit vector $\widehat e$ denoting direction of the light ray, which lets us write the Maupertuis action 
for light-like null curves:
\begin{equation}
\label{repa}\frac{d \ }{d \sigma} = \frac1{|\dot{\vec r}|} \frac{d \ }{d \tau} \qquad \Rightarrow \qquad \widehat{e} = \frac{d {\vec r}}{d \sigma} = \frac{\dot{\vec r}}{|\dot{\vec r}|} , \qquad \qquad |\widehat e|^2 = 1,
\end{equation}
$${\vec p}_{null} = \frac{\partial L_{null}}{\partial \dot{\vec r}} = n ({\vec r}) {\widehat e} \qquad \Rightarrow \qquad l = \int_1^2 \vec p_{null} . d \vec r = \int_1^2 n ({\vec r}) | d {\vec r} |.$$
Since the null geodesic path integral is given as shown above, we can conclude as shown in \cite{cgg} 
that for the geodesic time integral $\mathcal T = \int_1^2 d \tau \ {\dot t} = \frac1c \int_1^2 dl$, which leads to the Eikonal equation
\begin{equation}
\label{eikonal} \frac{\partial \mathcal T}{\partial {\vec r}} = \frac1c \frac{\partial l}{\partial {\vec r}} = \frac{\vec p_{null}}c = \frac{n ({\vec r})}c {\widehat e} \qquad \Rightarrow \qquad \left| \frac{\partial \mathcal T}{\partial {\vec x}} \right|^2 = \frac{n^2}{c^2}.
\end{equation}
Furthermore, we can use (\ref{repa}) to rewrite the Euler-Lagrange equation (\ref{optlag}) and derive from it a result \cite{br} normally derived from the Eikonal equation:
\begin{equation}
\label{imp} \frac{d \ }{d \sigma} \left( n ({\vec r}) \widehat{e} \right) = {\vec \nabla} n ({\vec r}) \qquad \Rightarrow \qquad \frac{d {\widehat e}}{d \sigma} = \left( {\widehat e} \times {\vec \nabla} \ln n \right) \times {\widehat e}.
\end{equation}
These spatial geodesics that such equations describe are better analysed by using the orthonormal 
frame of the Frenet-Serret formalism \cite{bgj}. We will use this result to demonstrate an Snell's 
law for refractive optics is applicable to gravitational fields as well. \smallskip

Now choose a basis in two dimensions $\left( {\widehat e}_{\parallel}, {\widehat e}_{\perp} \right)$ set up around the direction of ${\vec \nabla} n ({\vec r})$, where ${\widehat e}_{\parallel}$ 
denotes direction along ${\vec \nabla} n ({\vec r})$, while ${\widehat e}_{\perp}$ denotes direction orthogonal to it. We can therefore 
write for unit vector and derivative of refractive index:
$${\widehat e} = \cos \theta \ {\widehat e}_{\parallel} + \sin \theta \ {\widehat e}_{\perp}, \qquad \qquad \frac{d n}{d \sigma} = \frac{d {\vec r}}{d \sigma} . {\vec \nabla} n = {\widehat e} . {\vec \nabla} n = | {\vec \nabla} n | \cos \theta.$$
Applying the above equations to (\ref{imp}) gives us the LHS and RHS as follows, resulting in a 
solvable differential equation:
$$\frac{d {\widehat e}}{d \sigma} = \left( - \sin \theta \ {\widehat e}_{\parallel} + \cos \theta \ {\widehat e}_{\perp} \right) \frac{d \theta}{d \sigma}, \qquad \qquad \left( {\widehat e} \times {\vec \nabla} \ln n \right) \times {\widehat e} = \sin \theta \left( \sin \theta \ {\widehat e}_{\parallel} - \cos \theta \ {\widehat e}_{\perp} \right) \frac{ | {\vec \nabla} n|}n$$
Thus we have the conserved quantity of null-geodesic dynamics
\begin{equation}
\label{snell} n ({\vec r}) \sin \theta = const.
\end{equation}
which is Snell's Law from refractive optics. This supports the theory that regions with gravitational 
fields can be regarded as refractive media, and vice versa. Now, we will describe and analyze classical particle mechanics in optical terms.

\numberwithin{equation}{section}

\section{Optical-mechanical formulation}

For those solutions where $AB = 1$, we will proceed to rewrite the metric (\ref{isotrop}) in as conventional a form as possible, and 
deduce the form of the non-relativistic Lagrangian. A metric that is a solution to Einstein's 
equations for vaccum ($AB = 1$) can be written as:
$$ds^2 = \frac{c^2 dt^2}{n ({\vec r})} - n ({\vec r}) |d {\vec r}|^2 \quad = \quad c^2 dt^2 \left[ \frac1{n ({\vec r})} - n ({\vec r}) \left( \frac{|{\vec v}|}c \right)^2 \right].$$
If $A ({\vec r}) = \dfrac1{n ({\vec r})}= 1 + \dfrac{2 U ({\vec r})}{mc^2}$, we can say that:
\begin{equation}
\label{oemet} ds^2 = c^2 dt^2 \left[ 1 - \frac2{mc^2} \left\{ \left( \frac12 m n ({\vec r}) |{\vec v}|^2 \right) - U ({\vec r}) \right\} \right].
\end{equation}
Starting from this metric, we shall describe the optical-mechanical formulation of classical 
mechanics on such spaces. Since $m n({\vec r})$ acts effectively as position-dependent mass, we can write the kinetic energy $T$, non-relativistic Lagrangian $L$, and the energy $E$ via Legendre's transformation as: 
\begin{equation}
\label{nrlag} \begin{split} 
T &= \frac12 m n({\vec r}) |{\vec v}|^2, \qquad \qquad L = T - U ({\vec r}), \\
{\vec p} &= \frac{\partial L}{\partial {\vec v}} \qquad \Rightarrow \qquad E = {\vec p} . {\vec v} - L = T + U.
\end{split}
\end{equation}
If we define the rest Lagrangian $L_0$ and rest energy $E_0$ as $L_0 = - E_0 = - mc^2$, then for $U = \frac{E_0}2 \left(\frac1{n({\vec r})} - 1 \right)$, according to \cite{cg}, we can write the optical-mechanical relativistic Lagrangian from (\ref{oemet}), using (\ref{nrlag}),  as follows:
$${\mathcal L} = - mc \left| \frac{d s}{d t} \right| = L_0 \sqrt{1 + 2 \frac{L}{L_0}}, \qquad \qquad L_0 = - mc^2.$$
In the classical limit $\left| L \right| << |L_0|$, we have:
$${\mathcal L} \xrightarrow{|L| << |L_0|} L = L_0 + L \quad \equiv \quad \frac m2 \left[ n ({\vec r}) |{\vec v}|^2 - \frac{c^2}{n ({\vec r})} \right] + \frac{L_0}2.$$
Omitting the additive constant $\frac{L_0}2$, the effective classical Lagrangian parametrized wrt $\tau$ is:

\begin{align}
\label{clag}  \quad L \approx - \frac m2 \left( \frac{d s}{d \tau} \right)^2 = \frac m2 n ({\vec r}) |\dot{\vec r}|^2 - \frac{E_0 {\dot t}^2}{2 n ({\vec r})} \quad \Rightarrow \quad 
\begin{cases}
{\vec p} = \quad \dfrac{\partial L}{\partial \dot{\vec r}} = m n ({\vec r}) \dot{\vec r} \smallskip \\
p_t = - \dfrac{\partial L}{\partial {\dot t}} = - \dfrac{E_0}{n ({\vec r})} {\dot t} = - E
\end{cases} .
\end{align}
which can produce non-relativistic equations of motion on curved space. Deducing this form 
of Lagrangian from the non-relativistic equations of motion will lead to the spacetime metric. 

From the Euler-Lagrange equation applied to the non-relativistic Lagrangian $L$ (\ref{nrlag}), we 
can write the non-relativistic equations of motion for this space as:

$$m \frac{d \ }{d t} \left( n ({\vec r}) \vec v \right) = \frac m2 \left( \vec{\nabla} n (\vec r) \right) \left| \vec v \right|^2 - {\vec \nabla} U, \qquad \qquad \vec v = \frac{|\dot r|}{\dot t} = \frac{d \vec r}{d t},$$
\begin{equation}\label{Eq1} 
\frac{d^2 \vec r}{d t^2} + \frac1{2 n({\vec r})} \left[ 2 \left( \frac{d \vec r}{d t} . {\vec \nabla}n ({\vec r}) \right) \frac{d \vec r}{d t} - \left( \vec{\nabla} n (\vec r) \right) \left| \frac{d \vec r}{d t} \right|^2 \right] + {\vec \nabla} \left( \frac{c^2}{4 \left( n({\vec r}) \right)^2} \right) = 0.
\end{equation}
This equation is comparable to a Gorringe-Leach mechanical 
system with damping quadratically dependent on velocity. This shows that a spacetime that 
is conformally flat in the spatial part will exhibit some form of viscous drag.

If we consider the Lagrangian (\ref{nrlag}) of the system in one dimension only ($\vec r \rightarrow x$), then we will have $\left( \vec v . {\vec \nabla}n ({\vec r}) \right) \vec v = \left( \vec{\nabla} n (\vec r) \right) \left| \vec v \right|^2 \equiv n' (\vec x) \dot x^2$,
\begin{equation}
\ddot x + \frac{n' (x) \dot x^2}{2 n(x)} + \frac{U' (x)}{n(x)} = 0, \qquad \qquad \text{where } \ f' (x) = \frac{d f}{d x}.
\end{equation}
which is a Li\'enard type equation studied in \cite{cgcg} in the context of Jacobi-Maupertuis description. 
Under a weak perturbation of spacetime $\left( \frac{U (\vec r)}{c^2} \approx 0 \right)$, we can write (\ref{Eq1}) as follows
\[ \begin{split}
\vec \nabla n (\vec r) = \frac2{mc^2} \vec \nabla U (\vec r) \approx 0 \\ 
{\vec \nabla} \left( \frac{c^2}{4 \left( n({\vec r}) \right)^2} \right) = \frac1{2 n (\vec r)} \vec \nabla \left( \frac{c^2}{n (\vec r)} \right) = \frac2m \vec \nabla U (\vec r)
\end{split} \qquad \Rightarrow \qquad 
\begin{split}
\frac{d^2 \vec r}{d t^2} + {\vec \nabla} U (\vec r) = 0.
\end{split} \]
returning us to the familiar equation of motion for Newtonian gravitational potentials.

%We shall define the rest Lagrangian $L_0$ and rest energy $E_0$, to better write the relativistic Lagrangian as:
%\begin{equation}
%\label{rlag} L_0 = \frac{E_0}2 = \frac{mc^2}2 \qquad \Rightarrow \qquad {\mathcal L} = - 2 L_0 \sqrt{1 - \frac{L}{L_0}}.
%\end{equation} 

\numberwithin{equation}{subsection}

\subsection{Jacobi-Maupertuis description}

In Maupertuis form, the geodesic action integral can be described using the Maupertuis 
Lagrangian $L_{Maup}$, written using (\ref{clag}), for which the overall geodesic Hamiltonian vanishes  \cite{cgg}
\begin{equation}
\label{maup} L_{Maup} = p_{\mu} {\dot x}^{\mu} \ = \ {\vec p} . \dot{\vec r} + p_t {\dot t} \quad \equiv \quad m n(\vec r) \left| \dot{\vec r} \right|^2 - \dfrac{E_0 \dot t^2}{n ({\vec r})} = 2 L,
\end{equation}
\begin{equation}
\label{nullham} \Rightarrow \qquad \mathcal{H} = p_{\mu} {\dot x}^{\mu} - L_{Maup} = \frac1{2m n ({\vec r})} |{\vec p}|^2 - \frac{n ({\vec r})}{2mc^2} E^2 = 0.
\end{equation}
This means that from (\ref{maup}), the classical Hamilton-Jacobi equation, using (\ref{nrlag}) we can get 
\begin{equation}
\label{mact} S = \int_1^2 d \tau \ L_{Maup} = \int_1^2 \left( {\vec p} . d {\vec r} + p_t dt \right) \qquad \Rightarrow \qquad \frac{\partial S}{\partial {\vec r}} = \frac{\partial L}{\partial \dot{\vec r}} = {\vec p} = m n ({\vec r}) \dot{\vec r}.
\end{equation}
\begin{equation}
\label{hamjac}
E = \frac{|{\vec p}|^2}{2 m n ({\vec r})} + U ({\vec r}) \qquad \Rightarrow \qquad  \left| \frac{\partial S}{\partial {\vec r}} \right|^2 = 2 m n ({\vec r}) \left( E - U ({\vec r}) \right).
\end{equation}
Furthermore, from (\ref{nullham}) and (\ref{hamjac}), we can say that:
\begin{equation}
\label{spec} |{\vec p}|^2 = 2 m n ({\vec r}) \left( E - U ({\vec r}) \right) = \left( \frac{n ({\vec r}) E}c \right)^2 \qquad \Rightarrow \qquad E - U ({\vec r}) = \frac{E^2 n ({\vec r})}{2mc^2}.
\end{equation}
which is in contrast with what is stated in \cite{alsing, clanc}, where 
$$\frac nc = \alpha \sqrt{2m \left( E - U \right)} \quad \Rightarrow \quad n \sim \sqrt{2m \left( E - U \right)}.$$
Now, since the metric is time-independent, 
the momentum conjugate to time is constant $p_t \approx - E$, so we can say that the effective 
action $S_{eff}$ is given from (\ref{mact}) \cite{cgg} by:
$$\delta S = \delta \int_1^2 dt \left( {\vec p} . {\vec v} - E \right) = \delta \int_1^2 dt \ {\vec p} . {\vec v} = \delta \int_1^2 dt \ 2 T = 0, \qquad \qquad {\vec p} = \frac{\partial L}{\partial {\vec v}} = m n ({\vec r}) {\vec v},$$
$$\Rightarrow \qquad S_{eff} = \int_1^2 d \tau \left| \frac{d s_{cJ}}{d \tau} \right| = \int_1^2 dt \ \sqrt{T} \sqrt{4 T} = \int_1^2 dt \ \sqrt{ E - U ({\vec r}) } \sqrt{2m n ({\vec r}) |{\vec v}|^2}.$$
showing that the effective action covers only the spatial part of the geodesic, and according to \cite{cgg}, using (\ref{spec}) the classical Jacobi metric is given by:
\begin{equation}
\label{classjac} ds^2_{c J} = 2 m n ({\vec r}) \left( E - U ({\vec r}) \right) | d {\vec r} |^2 = \left( \frac{E n ({\vec r})}c \right)^2 | d {\vec r} |^2.
\end{equation}
and the relativistic Jacobi metric according to \cite{gwg, cgg}, is given by:
\begin{equation}
\label{reljac} ds^2_{rJ} = \left( \frac{\mathcal E^2}{c^2} n ({\vec r}) - m^2 c^2 \right) n ({\vec r}) | d {\vec r} |^2.
\end{equation}
Now we shall elaborate on the geometric formulation of damped mechanical systems.

\subsection{Equations with drag terms and related geometry}

We have seen in (\ref{Eq1}) that motion through a spacetime described by the metric (\ref{oemet}) is 
influenced by drag. There is another mechanical system influenced by a different form of drag 
according to its equation of motion, known as the Gorringe Leach equation \cite{gbm}. We shall 
study it to see how this form of damping affects the optical properties of its corresponding 
spacetime metric. Such systems must be carefully examined as they may or may not be integrable. 
We shall investigate some integrable cases of mechanical systems with 
drag terms. \bigskip

\noindent
For the general form of a metric with only a scalar potential
\begin{equation}
\label{met} ds^2 = A ({\vec x}) c^2 d t^2 - | d {\vec x} |^2, \qquad \quad A ({\vec x}) = 1 + \frac{2 U ({\vec x})}{mc^2}.
\end{equation}
and its effective classical Lagrangian given by (\ref{clag}) $L = - \frac m2 \left( \frac{d s}{d t} \right)^2$, we shall get the following equation of motion, and refractive index according to (\ref{refindex}):
\begin{equation}
\label{classeom} {\vec x}'' = - \frac{c^2}2 {\vec \nabla} A ({\vec x}) \equiv - \frac1m {\vec \nabla} U, \qquad \qquad n (\vec x) = \frac1{\sqrt{A (\vec x)}}.
\end{equation}
Now consider a damped equation of motion different from (\ref{Eq1}) given by
\begin{equation}
\label{damp} \ddot{\vec x} + h(\vec{x}; \dot{\vec x}) \dot{\vec x} + \frac{c^2}2 {\vec \nabla} \left( 1 + \frac{2U}{mc^2} \right) = 0.
\end{equation}
Reparametrization $t \longrightarrow \tau = \tau(t)$ to convert (\ref{damp}) to a more suitable form gives:
\begin{equation}
\label{repara} {\vec x}'' + \frac{\ddot{\tau} + \dot{\tau} h}{\dot{\tau}^2} {\vec x}' + \frac 1{m \dot{\tau}^2} {\vec \nabla} U  = 0.
\end{equation}
We can choose the reparametrization such that:
\begin{equation} \label{para}
\ddot{\tau} + \dot{\tau} h = 0 \qquad \Rightarrow \qquad \dot{\tau} = \dot{\tau}_0 e^{- \int dt . h} = e^{- \alpha (t)}, \qquad \quad \dot{\tau}_0 = 1.
\end{equation}
thus transforming (\ref{repara}) into a more familiar form similar to (\ref{classeom}):
\begin{equation}
\label{repareom} {\vec x}'' = - \frac{c^2 \text{e}^{2 \alpha (t)}}2 {\vec \nabla} \left( 1 + \frac{2U}{mc^2} \right).
\end{equation}
Thus, upon comparing (\ref{repareom}) to (\ref{classeom}), we will have the corresponding classical Lagrangian:
$${\widetilde L} = \frac m2 \left( \big| {\vec x}' \big|^2 - \text{e}^{2 \alpha} c^2 A (\vec{x}) \right) = \text{e}^{2 \alpha} \frac m2 \left( | \dot{\vec x} |^2 - c^2 A (\vec{x}) \right).$$
then using (\ref{para}), we have $d \tau = \text{e}^{- \alpha({\vec x}, t)} d t$, and in the classical limit, we will have the classical 
action invariant under parametrization $S_{class}$, given as
\begin{equation}
\label{dmplag} \widetilde{L} \ d \tau = L \ dt = \text{e}^{- \alpha} {\widetilde L} \ dt \qquad \Rightarrow \qquad L = \text{e}^{- \alpha} {\widetilde L} = \text{e}^{\alpha} \frac m2 \left( | \dot{\vec x} |^2 - c^2 A (\vec{x}) \right).
\end{equation}
Thus, the damped Lagrangian can be written by comparing (\ref{classeom}) to (\ref{repareom}), and the metric producing (\ref{damp}) just as (\ref{met}) produces (\ref{classeom}) will be according to (\ref{clag}):
\begin{equation}
\label{dmpmet} ds^2 = - \frac2m L \ dt^2 = \text{e}^{\alpha(t)} \left[ A (\vec x) c^2 d t^2 - |d {\vec x}|^2 \right].
\end{equation}
where the null geodesics of the isotropic metric (\ref{dmpmet}) have the isotropic refractive index $n (\vec x) = \left( \sqrt{A (\vec x)} \right)^{-1}$, the same as in (\ref{classeom}). \bigskip

Thus, we have obtained the form of the metric and the 
Lagrangian with damping effects applied by converting it into an undamped form via suitable 
reparametrization. Conversely, undamped systems can also be described as damped systems 
via suitable reparametrization. Furthermore, although the metric and Lagrangian obtained were time dependent, the equivalent refractive index was not.

\subsubsection{The Gorringe-Leach equation}

In 1993, Gorringe and Leach \cite{gbm} exhibited two classes of differential equations incorporating 
drag terms while having closed elliptical orbits
\begin{equation}
\label{gleach} \ddot{z} + h(z, \bar{z}; \dot{z}, \dot{\bar{z}}) \dot{z} + g(z, \bar{z}) z = 0.
\end{equation}
We transform (\ref{gleach}) into the form of (\ref{repareom}), which for a damped spherically symmetric Harmonic Oscillator is:
$$z'' = - g(z, \bar{z}) e^{2 \alpha (t)} z = - \omega^2 z.$$
If we write $H(z, \bar{z}) = \alpha (t)$, then we have:
\begin{equation}
\label{pot} g(z, \bar{z}) e^{2\alpha(t)} z = e^{2\alpha(t)} \frac{\partial U(|z|)}{\partial {\bar z}} = \omega^2 z \qquad \Rightarrow \qquad U (|z|) = e^{-2\alpha(t)} \frac{\omega^2}2 |z|^2.
\end{equation}
Working backwards from the Euler-Lagrange equation of motion and using (\ref{pot}), we can 
say that according to (\ref{dmplag}), we can write the Lagrangian as in \cite{cdgh}:
$$ L = \frac m2 \text{e}^{- \alpha(t)} \left[ | \dot{z} |^2 - c^2 \left( 1 + \frac{\omega^2 |z|^2}{c^2} \right) \right].$$
and the metric and isotropic refractive index for (\ref{gleach}) according to (\ref{dmpmet}) is:
$$ds^2 =  \text{e}^{- \alpha(t)} \left[ \frac{c^2 d t^2}{\left\{ n (|z|) \right\}^2} - |dz|^2 \right], \qquad \qquad \left( n (|z|) \right)^{-2} = 1 + \frac{\omega^2 |z|^2}{c^2}.$$
Thus, we have the metric and the Lagrangian for the Gorringe Leach equation. If we were to write (\ref{Eq1}) in a plane in complex variables, then we would have:
$$\vec r \ \rightarrow \ z, \qquad \qquad \text{where } \ \vec r = x \ \widehat e_x + y \ \widehat e_y, \quad \text{and } \quad z = x + i y,$$
\begin{equation}
\label{spsq} \left|\frac{d \vec r}{d t} \right|^2 = \frac{d \bar z}{d t} . \frac{d z}{d t} = \left| \frac{d z}{d t} \right|^2,
\end{equation}
$$\vec{\nabla} n (\vec r) = \widehat e_x \ \frac{\partial n}{\partial x} + \widehat e_y \ \frac{\partial n}{\partial y} \ \equiv \ \frac{\partial n}{\partial x} + i \ \frac{\partial n}{\partial y} = \frac{\partial z}{\partial x} \frac{\partial n}{\partial z} + \frac{\partial \bar z}{\partial x} \frac{\partial n}{\partial \bar z} + i \left( \frac{\partial z}{\partial y} \frac{\partial n}{\partial z} + \frac{\partial \bar z}{\partial y} \frac{\partial n}{\partial \bar z} \right),$$
\begin{equation}
\label{cplxgrad} \Rightarrow \qquad \vec{\nabla} n (\vec r) \longrightarrow \nabla_z n = \frac{\partial n}{\partial z} + \frac{\partial n}{\partial \bar z} + i \left( i \frac{\partial n}{\partial z} - i \frac{\partial n}{\partial \bar z} \right) = 2 \frac{\partial n}{\partial \bar z}.
\end{equation}
Thus, according to (\ref{cplxgrad}), we can say that:
\begin{equation}
\label{tderiv} \frac{d n}{d t} = \frac{d \vec r}{d t} . \vec{\nabla} n (\vec r) = \frac12 \left( \frac{d z}{d t} \nabla_{\bar z} n + \frac{d \bar z}{d t} . \nabla_z n \right) = \frac{d z}{d t} \frac{\partial n}{\partial z} + \frac{d \bar z}{d t} \frac{\partial n}{\partial \bar z}.
\end{equation}
Therefore, using (\ref{spsq}), (\ref{cplxgrad}) and (\ref{tderiv}) we can write (\ref{Eq1}) as follows:
$$\frac{d^2 z}{d t^2} + \frac1n \left[ \frac{d n}{d t} \frac{d z}{d t} - \frac{\partial n}{\partial \bar z} \left|\frac{d z}{d t} \right|^2 \right] + \frac{\partial \ }{\partial \bar z} \left( \frac{c^2}{2 n^2} \right) = 0.$$
which can be further re-written as shown below with the following replacements:
\begin{equation}
\label{finaleq} \frac{d^2 z}{d t^2} + \frac1n \left[ \frac{d n}{d t} - \frac{\partial n}{\partial \bar z} \frac{d \bar z}{d t} \right] \frac{d z}{d t} + \left[ \frac1z \frac{\partial \ }{\partial \bar z} \left( \frac{c^2}{2 n^2} \right) \right] z = 0.
\end{equation}
$$h(z, \bar z; \dot z, \dot{\bar z}) = \frac1n \left[ \frac{d n}{d t} - \frac{\partial n}{\partial \bar z} \frac{d \bar z}{d t} \right], \qquad \text{and } \qquad g(z, \bar z) = \frac1z \frac{\partial \ }{\partial \bar z} \left( \frac{c^2}{2 n^2} \right).$$
Thus, we can clearly see that (\ref{finaleq}) has fit into the form of (\ref{gleach}), showing that it is truly comparable to a Gorringe-Leach equation.

\subsubsection{Damped Kepler-Hooke duality}

Using (\ref{para}), (\ref{gleach}) and (\ref{pot}) will lead to the conserved Fradkin tensor:
\begin{equation}
\label{frad} \ddot{z} + \dot{\alpha} \dot{z} + \omega^2 e^{-2\alpha(t)} z = 0 \qquad \Rightarrow \qquad \mathcal{J}_{zz} = \dot{z}^2 e^{2\alpha(t)} + \omega^2 z^2.
\end{equation}
Applying the Bohlin map $\xi = z^2$ and re-parametrization $d \sigma = |z|^2 d t = |\xi| d t$ rewrites the 
Fradkin tensor $\mathcal{J}_{zz}$ (\ref{frad}) into:
\begin{equation}
\label{frad1} \mathcal{J}_{zz} = \frac 1 4 \bar{\xi} \left( \xi' \right)^2 e^{2 \alpha(t)} + \omega^2 \xi, \qquad \qquad \text{where } \ \xi' = \frac{d \xi}{d \sigma} = \frac{2 z}{|z|^2}  \dot z.
\end{equation}
Isotropy in a system implies that the equation of motion takes the same form along any direction axis. We can use the system  isotropy to infer the conjugate equation from (\ref{frad}):
\begin{equation}
\label{gleach2} \ddot{\bar{z}} + \dot{\alpha} \dot{\bar{z}} + \omega^2 e^{-2\alpha(t)} \bar{z} = 0.
\end{equation}
Thus, using the equations from (\ref{frad}) and (\ref{gleach2}), we get the conserved quantity $\mathcal{J}_{z \bar{z}}$ which transforms under the Bohlin map:
\begin{equation}
\label{frad2} \mathcal{J}_{z \bar{z}} = \vert \dot{z} \vert^2 e^ {2 \alpha(t)} + \omega^2 \vert z \vert^2 = \left( \frac 1 4 \vert \xi' \vert^2 e^{2 \alpha(t)} + \omega^2 \right) \vert \xi \vert \qquad \Rightarrow \quad \omega^2 = \frac{\mathcal{J}_{z\bar{z}}}{\vert \xi \vert} - \frac14 \vert \xi' \vert^2 e^{2 \alpha(t)}.
\end{equation}
Thus, using (\ref{frad1}) and (\ref{frad2}) gives us the result:
\begin{equation}
\label{rlenz} \mathcal{J}_{zz} = \frac { e^{2 \alpha(t)}} 4 \left( \bar{\xi} \xi' - \bar{\xi}' \xi \right) \xi' + \mathcal{J}_{z\bar{z}} \frac{\xi}{\vert \xi \vert}.
\end{equation}
Comparing (\ref{rlenz}) to $\mathbb{A} = -iL(m\xi') - 4m^2\mathcal{J}_{z \bar{z}} \frac{\xi}{\vert \xi \vert}$, where $L = - i m \left( \bar{\xi} \xi' - \bar{\xi}' \xi \right)$, we get the 
equivalent Runge-Lenz vector
\begin{equation}
\mathbb{A} = - 4 m^2 \mathcal J_{zz} = -iL(m\xi') e^{2 \alpha (t)} - 4m^2\mathcal{J}_{z \bar{z}} \frac{\xi}{\vert \xi \vert}.
\end{equation}
This system, appears to be a re-parameterized version of the original harmonic oscillator. 
Aside from the exponential factor, the form of the equivalent Runge-Lenz vector is the same.

\subsubsection{Damped Hamiltonian mechanics}

The classical Lagrangian for a dissipative system according to (\ref{dmplag}) can be given by:
\begin{equation}
\label{disslag} L = \text{e}^{\alpha(t)} \left( \frac m2 | \dot{\vec x} |^2 - U \right).
\end{equation}
The Hamiltonian by Legendre transformation of (\ref{disslag}) is:
\begin{equation}
\label{dissham} H = \frac1{2m} |{\vec p}|^2 \text{e}^{- \alpha (t)} + U (\vec{x}) \text{e}^{\alpha(t)}, \qquad \qquad {\vec p} = m \text{e}^{\alpha (t)} \dot{\vec x}.
\end{equation}
The Hamilton's equations of motion are given by:
\begin{equation}
\label{hamil} \dot{\vec x} = \frac{\partial H}{\partial {\vec p}} = \frac{\vec p}m \text{e}^{- \alpha (t)}, \qquad \qquad \dot{\vec p} = - \frac{\partial H}{\partial {\vec x}} = - {\vec \nabla} U \text{e}^{\alpha (t)}.
\end{equation}
Thus, we will find that the Hamiltonian (\ref{dissham}) is dissipative:
$$\frac{d H}{d t} = \left( \dot{\vec p} . \frac{\vec p}m \text{e}^{- \alpha (t)} + \dot{\vec x} . {\vec \nabla} U \text{e}^{\alpha (t)} \right) - \dot{\alpha} \left( \frac1{2m} |{\vec p}|^2 \text{e}^{- \alpha (t)} - U (\vec{x}) \text{e}^{\alpha(t)} \right) = - L \dot{\alpha}.$$
This concludes our optical analysis of mechanics and spacetime with drag included. As we 
can see, the central force, and consequently the potential are time dependent. Now we shall 
elaborate on the formulation of dynamics related to null geodesics.

\numberwithin{equation}{section}

\section{Dynamical solutions for null geodesics via Binet's equation}

In \hyperref[sec:sec2]{Sec. 2}, we deduced optical mechanics from null geodesics. Here, we shall study how Binet's 
equation derive from null geodesics and dynamically compare them to mechanical systems 
with a central force. Casey \cite{casey1} studied the Schwarzschild-Tangherlini metric, and here we have attempted to extend his work to other solutions. A spherically symmetric Lorentzian $(n+1)$-dimensional metric with 
$S^{n-1}$ symmetry that is asymptotically flat can be written as:
\begin{equation} \label{spmet} 
\begin{split} 
ds^2 = f(r) c^2 dt^2 - \frac{dr^2}{g(r)} - r^2 d \Omega_{n-1}^2 \qquad \text{where }
\end{split} \quad 
\begin{split}
f(r) = 1 + F(r), \quad F(r) = \sum_{i=2}^{\infty} a_i r^{-i} \\ 
g(r) = 1 + G(r), \quad G(r) = \sum_{i=2}^{\infty} b_i r^{-i}
\end{split}.
\end{equation}
Upon restriction to motion in the plane $\dot{\theta} = 0$ for $n = 2$, the Lagrangian according to (\ref{clag}) is:
\begin{equation}
\label{lag} L = \frac m2 \left( \frac{\dot{r}^2}{g(r)} + r^2 \dot{\phi}^2 - f(r) c^2 \dot{t}^2 \right).
\end{equation}
We should keep in mind that from \ref{lag}, we can deduce 2 conserved quantities:
\begin{equation}
\label{consv} q = - \frac1c \frac{\partial L}{\partial \dot{t}} = f(r) c \dot{t} \qquad \qquad l = \frac{\partial L}{\partial \dot{\phi}} = r^2 \dot{\phi}.
\end{equation}
The null geodesic is characterized by setting $ds^2 = 0 \ \Rightarrow \ L = 0$ for (\ref{spmet}) and (\ref{lag}). To provide 
the same formulation employed in \hyperref[sec:sec2]{Sec. 2}, we will define two null geodesics under constraints 
since the space is not isotropic to define the directional refractive indices according to (\ref{nisoindex}).
\begin{equation} \label{newri}
\begin{split}
\phi = constant \qquad & \frac{\dot{r}^2}{g(r)} = f(r) c^2 \dot{t}^2 \qquad \Rightarrow \qquad  n_r^2 = \frac1{f(r) g(r)} \\
r = constant \qquad & r^2 \dot{\phi}^2 = f(r) c^2 \dot{t}^2 \qquad \Rightarrow \qquad  n_\phi^2 = \frac1{f(r)}
\end{split} .
\end{equation}

Since we are dealing with null geodesics, the geodesic cannot be parametrised along a 
vanishing curve. However, we have seen that the geodesics can be produced by extremising the 
spatial curve as shown in (\ref{optlag}) to take the least time to traverse in accordance with Fermat's 
principle, with respect to which it can be parametrised. \bigskip

\noindent
Writing $ds_{\mathcal{O}}^2 = dt^2$ in (\ref{spmet}) for $n = 2$, describes the unparameterised geodesics of the 
optical 2-metric, which we can choose to write in isotropic co-ordinates:
\begin{equation}
\label{unpara} ds_{\mathcal{O}}^2 = \frac{dr^2}{f(r) g(r)} + \frac{r^2}{f(r)} d \phi^2 \quad \equiv \quad \left[ \eta (\rho) \right]^2 \left( d \rho^2 + \rho^2 d \phi^2 \right).
\end{equation}
\[ \begin{split} 
\eta (\rho) \ d \rho = \frac{dr}{\sqrt{f(r) g(r)}} \\ 
\eta (\rho) \rho = \frac{r}{\sqrt{f(r)}} 
\end{split} 
\qquad \Rightarrow \qquad 
\begin{split}
\frac{d \rho}{\rho} = \frac{dr}{r \sqrt{g(r)}}.
\end{split} \]
where the conformal factor $\eta (r)$ is the isotropic refractive index. In case of Schwarzschild 
solution with $n = 3$ and $f(r) = g(r) = 1 - \frac{2 M_3}r$ for motion in a plane, the isotropic co-ordinate 
is given by:
$$\frac{d \rho}{\rho} = \frac{dr}{\sqrt{r \left( r - 2 M_3 \right)}} \qquad \Rightarrow \qquad \rho = c \left( \frac r{M_3} - 1 + \frac1{M_3} \sqrt{ r \left( r - 2M_3 \right)} \right).$$
where $\rho = M_3$ for $r = 2 M_3$, making $c = M_3$. Thus, the isotropic co-ordinate $\rho$, and the 
conformal factor $\eta (\rho)$ are:
\begin{equation}
\label{iso1} \rho = r - M_3 + \sqrt{ r \left( r - 2M_3 \right)}, \qquad \qquad \eta (\rho) = \frac{\left( \rho + M_3 \right)^3}{2 \rho^2 \left( \rho - M_3 \right)}.
\end{equation}
We shall now discuss solutions to the null-geodesic equations that will help describe blackhole 
optics, and apply the formulation to other examples after demonstrating on one example 
discussed by Casey \cite{casey1}.

\numberwithin{equation}{subsection}

\subsection{Central force mechanics and Binet's equation}

We can see that the metric (\ref{spmet}) is spherically symmetric, which means that we are dealing with mechanical systems governed by central forces. For various force laws, the solutions to the equations will describe their respective orbits, 
allowing us to use existing solutions from dynamics to describe null-geodesic mechanics.  \bigskip \\
Non-relativistic central-force motion can be deduced from null geodesics of (\ref{spmet}). If we start with the null geodesic of (\ref{spmet}) and substitute the conserved quantities listed in (\ref{consv}) accordingly, we have the equation:
$$ds^2 = 0 \qquad \Rightarrow \qquad \frac{q^2}{f(r)} - \frac{\dot r^2}{g(r)} - \frac{l^2}{r^2} = 0 \qquad \Rightarrow \qquad \dot r^2 =  - l^2 \frac{g(r)}{r^2} + q^2 \frac{g(r)}{f (r)},$$
$$\Rightarrow \qquad \dot r^2 + \frac{l^2}{r^2} =  - l^2 \frac{G(r)}{r^2} + q^2 \frac{g(r)}{f (r)}$$
Taking the time derivative of the above equation, with $( \ )' = \frac{d \ }{d r}$ here, we get:
$$2 \dot r \left[ \ddot r - \frac{l^2}{r^3} \right] = \dot r \left[ - l^2 \left( \frac{G(r)}{r^2} \right)' + q^2 \left( \frac{g(r)}{f(r)} \right)' \right],$$
\begin{equation}
\label{eom} \Rightarrow \qquad \ddot{r} - r \dot{\phi}^2 = - \frac{l^2}2 \left( \frac{G(r)}{r^2} \right)' + \frac{q^2}2 \left( \frac{g(r)}{f(r)} \right)'.
\end{equation}
in a central force $F(r)$ for a potential $V(r)$ described by:
\begin{equation} \label{pots}
F(r) = - \frac{l^2}2 \left( \frac{G(r)}{r^2} \right)' + \frac{q^2}2 \left( \frac{g(r)}{f(r)} \right)', \qquad \qquad 
V(r) = \frac{l^2}2 \left( \frac{G(r)}{r^2} \right) - \frac{q^2}2 \left( \frac{g(r)}{f(r)} \right).
\end{equation}
This allows us to write the central force and potential for solutions to the Einstein's equations. The radial co-ordinate inversion $r \longrightarrow u = \frac1r$ for $L = 0$ in (\ref{lag}) gives the differential equation:
\begin{equation}
\label{mde1} \left( \frac{d u}{d \phi} \right)^2 + \widetilde{g}(u) u^2 = \left( \frac ql \right)^2 \frac{\widetilde{g}(u)}{\widetilde{f}(u)}, \qquad \qquad \text{where } \ \widetilde g (u) = g (r).
\end{equation}
In terms of (\ref{newri}), if we write $b = \frac lq$, then (\ref{mde1}) becomes:
\begin{equation} 
\label{mde2} \left( \frac{d u}{d \phi} \right)^2 + u^2 = \left[ 1 - \left( \frac{\widetilde n_\phi}{\widetilde n_r} \right)^2 \right] u^2 + \frac1{b^2} \left( \frac{\widetilde n_\phi^2}{\widetilde n_r} \right)^2.
\end{equation}
The corresponding Binet's equation can be deduced by differentiating the above equation:
\begin{equation}
\label{binet} \frac{d^2 u}{d \phi^2} + u = \frac{F'(u)}2, \qquad \qquad F(u) = \frac1{b^2} \left( \frac{\widetilde n_\phi^2}{\widetilde n_r} \right)^2 - \left[ 1 - \left( \frac{\widetilde n_\phi}{\widetilde n_r} \right)^2 \right] u^2.
\end{equation}
If we choose our co-ordinates such the spatial part is conformally flat like (\ref{isotrop}), ie.
$$ds^2 = f(r) c^2 dt^2 - \frac1{g(r)} \left( dr^2 + r^2 d \Omega_{n-1}^2 \right) \qquad \Rightarrow \qquad n_r^2 = n_\phi^2 = \frac1{f(r) g(r)} = n^2.$$
then we will instead get the equations equivalent to (\ref{mde2}) and (\ref{binet}) as:
\begin{align}
\label{mde2a} \left( \frac{d u}{d \phi} \right)^2 + u^2 &= \left( \frac{n (u)}b \right)^2, \\
\label{binet2} \frac{d^2 u}{d \phi^2} + u &= \frac{n (u) \ n'(u)}{b^2}.
\end{align}
Thus, solutions to Binet's equations available in dynamics should help describe the null-geodesic trajectories for various force laws. 

\subsection{Solutions and Schwarzschild Tangherlini metric}

Casey studied Optical metrics via null geodesics of the Schwarzschild-Tangherlini solution \cite{casey1}. 
Here we shall reproduce the Binet's equation deduced by Casey, derive the Schwarzschild-
Tangherlini solution, and write its optical and dynamical properties. In the next section, we 
shall extend his results to other solutions. \bigskip

\noindent
If we choose the following settings for $f(r) \equiv \widetilde f (u)$ and $g(r) \equiv \widetilde g (u)$ of (\ref{spmet}):
\begin{equation} \label{set}
\begin{split}
\widetilde g(r) &= 1 + A u^{n-2}, \qquad \qquad \frac{\widetilde{g}(u)}{\widetilde{f}(u)} = B u^n + C, \\
\lim_{u \rightarrow 0} \widetilde f (u) = \lim_{u \rightarrow 0} \widetilde g (u) &= 1 \qquad \Rightarrow \qquad C = 1 \qquad \Rightarrow \qquad \widetilde{f}(u) = \frac{1 + A u^{n-2}}{1 + B u^n}.
\end{split}
\end{equation}
the differential equation (\ref{mde2}) and (\ref{binet}) will become:
\begin{align}
\label{mde3} \left( \frac{d u}{d \phi} \right)^2 + u^2 &= 2 M_n u^n + \frac1{b^2} \qquad 2 M_n = - \left( A + \frac B{b^2} \right), \\
\label{binet3} & \frac{d^2 u}{d \phi^2} + u = n M_n u^{n-1} .
\end{align}
where (\ref{binet3}) is known as Binet's equation. If $B = 0$, we will have $2 M_n = - A$, meaning that according to (\ref{newri}) and (\ref{set}):
\begin{equation}
\label{schwri} f(r) = g(r) = 1 - \frac{2M_n}{r^{n-2}} \qquad \Rightarrow \qquad n_r (r) = \left( n_\phi (r) \right)^2 = \frac1{f(r)} = \left( 1 - \frac{2M_n}{r^{n-2}} \right)^{-1}.
\end{equation}
which results in the Schwarzschild-Tangherlini solution studied by Casey in \cite{casey1}:
\begin{equation}
\label{schtan} ds^2 = - \left( 1 - \frac{2M_n}{r^{n-2}} \right) dt^2 + \frac{dr^2}{1 - \dfrac{2M_n}{r^{n-2}}} + r^2 \ d \Omega_{n-2}^2.
\end{equation}
where according to (\ref{pots}), the central force and potential are:
\begin{equation}
F_{ST} (r) =  - \dfrac{n l^2 M_n}{r^{n+1}}, \qquad \qquad V_{ST} (r) = - l^2 \left( \dfrac{M_n}{r^n} \right) - \frac{q^2}2.
\end{equation}
Now we will look at two other solutions of Einstein's equations.

\subsection{Solutions of Binet's equations}

Using Binet's equation, it should be possible to deduce the deflection angle by solving for $\Delta \phi$ via integration. So, if we have the following equation from (\ref{mde3})
$$\left( \frac{d u}{d \phi} \right)^2 = \frac1{b^2} - u^2 + 2 M_n u^n,$$
then the closest proximity to the source along the null-geodesic trajectory is given by:
\begin{equation}
\label{umax} \left( \frac{d u}{d \phi} \right)_{u = u_m}^2 = \frac1{b^2} - u_m^2 + 2 M_n u_m^n = 0
\end{equation}
and using (\ref{umax}), the overall deflection angle is given by the definite integral:
\begin{equation}
\label{deflect} \mathcal D = \Delta \phi - \pi = 2 \int_0^{u_m} \frac{b \ d u}{\sqrt{ 1 - b^2 \left( u^2 - 2 M_n u^n \right)}} - \pi.
\end{equation}
If we define a new variable $v = \frac u{u_m}$, then we can write the integral (\ref{deflect}) as:
$$\Delta \phi = \frac2{\sqrt{K}} \int_0^1 \frac{d v}{\sqrt{ 1 - K^{-1} \left(v^2 - A_n v^n \right)}}, \qquad \quad \text{where } \quad A_n = 2 M_n u_m^{n-2}, \quad K = \frac1{b^2 u_m^2}.$$
Since $0 \leq v \leq 1$, we can expand the integrand binomially as follows:
$$\left[1 - K^{-1} \left(v^2 - A_n v^n \right) \right]^{-\frac12} \approx 1 + \frac1{2 K} \left(v^2 - A_n v^n \right) + \mathcal O (2).$$
So upto 1st order, we shall have:
$$\Delta \phi = \frac2{\sqrt{K}} \left[ 1 + \frac1{2 K} \int_0^1 dv \ v^2 \left( 1 - A_n v^{n-2} \right) \right],$$

\smallskip

\noindent
The Chebyshev theorem \cite{Chebyshev} integrals on differential binomials
$$ I = \int x^m(a + bx^n)^p \,dx $$ can be evaluated in terms of elementary functions if and only if

\begin{enumerate}
\item[(a)] $p$ is an integer, then we expand $(a + bx^n )^p$ by the binomial formula in order to rewrite the
integrand as a rational function of simple radicals $x^{j/k}$. By a simple substitution $x=t^r$ we
remove the radicals entirely and obtain integral on
rational function.
\item[(b)] $m+1/n$ is an integer, then setting $t = a + bx^n$ we convert the integral to $\int t^p (t-a)^{m=1/n - 1}dt$. 
\item[(c)] $m+1/n + p$ is an integer, then we transform the integral by factoring out $x^n$ and resultant
new integral of the differential binomial belongs to case (b).
\end{enumerate}

\smallskip

According to Chebyshev's theorem \cite{cgly, mz, plmath}, the solutions to the following indefinite integral in terms of incomplete beta function is:
\begin{equation}
\label{chebyshev} \int dx \ x^p \left( \alpha + \beta x^r \right)^q \ = \ \frac1r \alpha^{\frac{p+1}r + q} \beta^{-\frac{p+1}r} B_y \left( \frac{p+1}r, q - 1 \right), \qquad \quad y = \frac{\beta}{\alpha} x^r.
\end{equation}
So for partial deflections given by indefinite integrals, using (\ref{chebyshev}) we shall have
$$\phi = \int_0^u \frac{b \ d x}{\sqrt{ 1 - b^2 \left( x^2 - 2 M_n x^n \right)}} \approx \frac1{\sqrt K} \left[ x + \frac1{2 K} \int_0^x dv \ v^2 \left( 1 - A_n v^{n-2} \right) \right], \qquad u < u_m,$$
\begin{equation}
\label{partdefl} \therefore \qquad \phi = \frac1{\sqrt K} \left[ x + \frac{(- A_n)^{- \frac3{n-2}}}{2 K (n-2)} B_y \left( \frac3{n-2}, 0 \right) \right].
\end{equation}
Thus, for $n = 3$, we have the following solution given by indefinite integral:
\begin{equation}
\label{n=3} \Delta \phi_{n=3} (u) = \frac{2 (u - \beta_3) \sqrt{ \left( \frac{u - \beta_1}{\beta_3 - \beta_1} \right) \left( \frac{u - \beta_2}{\beta_3 - \beta_2} \right)} Elliptic F \left[ ArcSin \left\{ \sqrt{\frac{\beta_3 - u}{\beta_3 - \beta_2}} \right\}, \frac{\beta_2 - \beta_3}{\beta_1 - \beta_3} \right]}{\sqrt{\frac1{b^2} + u^2 (2 M_n u - 1)} \sqrt{\frac{u - \beta_3}{\beta_2 - \beta_3}}} 
\end{equation}

\begin{figure}
\begin{center}
\includegraphics[width=0.4\linewidth]{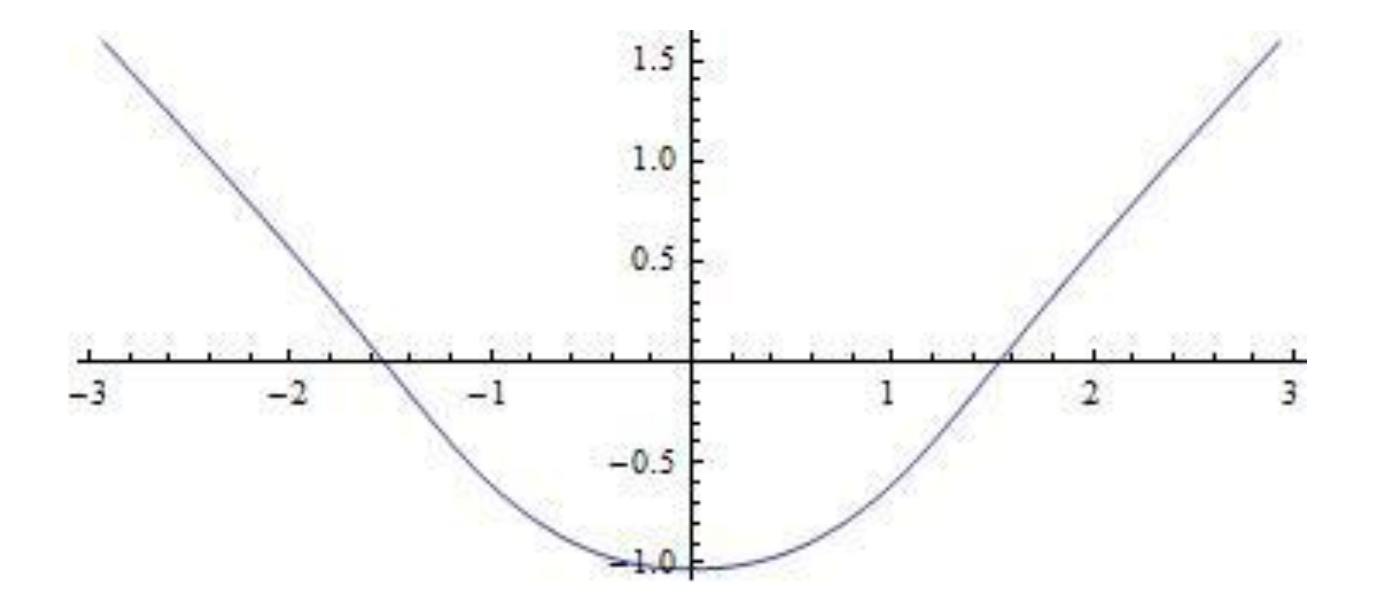}
\caption{\small Null-geodesic trajectory for $n = 3, b = \sqrt{3}, M_3 = \frac13$.}
\end{center}
\end{figure}

where $\beta_1, \beta_2, \beta_3$ are the roots of the equation $A_3 x^3 - x^2 + K = 0$. We shall now examine the Binet's equations formulated for other metrics.

\subsubsection{Helmholtz oscillators}

Most solutions of Einstein's equations for spherically symmetric spaces will have $f(r) = g(r)$. 
Here we have considered two such examples:
\begin{align}
f_H (r) &= 1 - \Lambda r^2 - \frac{2M}r = 1 - \frac{\Lambda}{u^2} - 2Mu, \\
f_{HD} (r) &= 1 - \Lambda r^2 - \frac Q{r^2} - \frac{2M}r = 1 - \frac{\Lambda}{u^2} - Qu^2 - 2Mu.
\end{align}
According to (\ref{pots}), the central forces and potentials for these solutions respectively are:
\begin{eqnarray}
&F_H (r) = - \dfrac{3 l^2 m}{r^4}, \qquad \qquad &V_H (r) = - l^2 \left[ \frac{\Lambda}2 + \left( \dfrac{m}{r^3} \right) \right] - \frac{q^2}2, \\
&F_{HD} (r) =  - l^2 \left[ \dfrac{3 m}{r^4} + \dfrac{2 Q}{r^5} \right], \qquad \qquad &V_{HD} (r) = - l^2 \left[ \frac{\Lambda}2 + \left( \dfrac m{r^3} \right) + \dfrac Q{2 r^4} \right] - \frac{q^2}2.
\end{eqnarray}
The refractive indices of these spaces are given as they were in (\ref{schwri}):
\begin{align}
n_{rH} (r) &= \left( n_{\phi H} (r) \right)^2 = \left( 1 - \frac{\Lambda}{u^2} - 2Mu \right)^{-1}, \\
n_{rHD} (r) &= \left( n_{\phi HD} (r) \right)^2 = \left( 1 - \frac{\Lambda}{u^2} - Qu^2 - 2Mu \right)^{-1}.
\end{align}
The related corresponding Binet's equations are given respectively as
\begin{align}
\label{hmhtz} \frac{d^2 u}{d \phi^2} + u &= 3Mu^2, \\
\label{helmduff} \frac{d^2 u}{d \phi^2} + u &= 3Mu^2 + 2Qu^3.
\end{align}
The above 2 results are equivalent to the equations for the Helmholtz oscillator \cite{almsan}, and the 
Helmholtz-Duffing oscillator \cite{asyyky} respectively, both of which are nonlinear equations that have 
received a lot of attention recently for the wide range of applications in engineering. The 
solution of the Helmholtz equation (\ref{hmhtz}) is given in terms of Jacobi elliptic function $sn$. The 
exact solution of the Helmholtz-Duffing oscillator equation (\ref{helmduff}) can also be expressed in 
terms of Jacobi elliptic function \cite{cvet,cvet1}. It should be noted that Gibbons and Vyska \cite{gwg1} used 
Weierstrass elliptic functions to give a full description and classification of null geodesics in 
Schwarzschild spacetime. 

\subsubsection{Kerr metric}

The black-hole spacetime known as the rotating (Kerr) black hole is a stationary metric.  \bigskip

\noindent
The Kerr metric (setting $c = 1$) is:
\begin{equation}
\begin{split}
d s^2 = \left( 1 - \frac{2 G M r}{\rho^2} \right) c^2  dt^2 &- \frac{4 G M a r \sin^2 \theta}{\rho^2} d \phi \ c \ d t - \frac{\rho^2}{\Delta} dr^2 \\
& \quad - \rho^2 \ d \theta^2 - \frac{\sin^2 \theta}{\rho^2} \left[ \left( r^2 + a^2 \right)^2 - a^2 \Delta \sin^2 \theta \right] d \phi^2, \\ \\
\Delta (r) = r^2 &- 2 G M r + a^2 \qquad \rho^2 (r, \theta) = r^2 + a^2 \cos^2 \theta.
\end{split}
\end{equation}
Using the formulation for non-isotropic spaces (\ref{niso}) - (\ref{nisoindex}), we have the following refractive indices:
\begin{equation}
\begin{split}
n_r &= \frac{\rho^2}{\sqrt{\Delta \left( \rho^2 - 2 G M r \right)}}, \qquad \qquad n_\theta = \frac{\rho^2}{\sqrt{\rho^2 - 2 G M r}}, \bigskip \\ 
n_\phi &= \frac{4 G M a r \sin^2 \theta}{\rho^2 - 2 G M r} + \sin \theta \sqrt{\left(\frac{2 G M a r \sin \theta}{\rho^2 - 2 G M r} \right)^2 + \frac{\left( r^2 + a^2 \right)^2 - a^2 \Delta \sin^2 \theta}{\rho^2 - 2 G M r}}.
\end{split}
\end{equation}
If $\theta = \frac{\pi}2$, then we will have $\rho^2 = r^2, \sin \theta = 1$, and
$$ds^2 = \left( 1 - \frac{2 G M}r \right) c^2 dt^2 - \frac{4 G M a}r c \ d t \ d \phi - \frac{r^2}{\Delta} dr^2 - \left( r^2 + a^2 + \frac{ 2 GM a^2}r \right) d \phi^2.$$
For the following classical Lagrangian, we have the conserved quantities:
$$L = - \frac12 \left( \frac{ds}{dt} \right)^2 = \frac12 \left[ \frac{r^2}{\Delta} \dot r^2 + \left( r^2 + a^2 + \frac{ 2 GM a^2}r \right) \dot \phi^2 + \frac{4 G M a}r c \dot t \dot \phi - \left( 1 - \frac{2 G M}r \right) c^2 \dot t^2 \right],$$
\begin{equation} \label{kerrcons}
\begin{split}
q = - \frac1c \frac{\partial L}{\partial \dot t} \ &= \ \left( 1 - \frac{2 G M}r \right) c \dot t - \frac{2 G M a}r \dot \phi, \\
l = \frac{\partial L}{\partial \dot \phi} \ &= \ \frac{2 G M a}r c \dot t + \left( r^2 + a^2 + \frac{ 2 GM a^2}r \right) \dot \phi.
\end{split}
\end{equation}
From (\ref{kerrcons}), we can deduce that:
\[ \begin{split}
c \dot t &= \Omega \left[ \gamma q + \beta l \right], \\ 
\dot \phi &= \Omega \left[ \alpha l - \beta q \right]
\end{split} \quad \text{where } \quad 
\begin{split}
\alpha (r) &= 1 - \dfrac{2 GM}r, \quad \beta (r) = \dfrac{2 G M a}r, \quad \gamma (r) = r^2 + a^2 + \dfrac{2 GM a^2}r, \\ 
&\Omega (r) = \left[ \dfrac{ 2 GM a^2}r + (r^2 + a^2 ) \left(1 - \dfrac{2 GM}r \right) \right]^{-1}
\end{split} \]
So, for null geodesic $ds^2 = 0$, we will have:
\[ \begin{split}
\frac{r^2}{\Delta} \dot r^2 &= \left( 1 - \frac{2 G M}r \right) c^2 \dot t^2 - \frac{4 G M a}r c \dot t \dot \phi - \left( r^2 + a^2 + \frac{ 2 GM a^2}r \right) \dot \phi^2, \\
&= \Omega^2 \left[ \alpha \left( \gamma q + \beta l \right)^2 - 2 \beta \left( \gamma q + \beta l \right) \left( \alpha l - \beta q \right) - \gamma \left( \alpha l - \beta q \right)^2 \right],
\end{split} \]
$$\Rightarrow \qquad \dot r^2 = \left( \alpha (r) + \frac{a^2}{r^2} \right) \Omega^2 \left[ \alpha \left( \gamma q + \beta l \right)^2 - 2 \beta \left( \gamma q + \beta l \right) \left( \alpha l - \beta q \right) - \gamma \left( \alpha l - \beta q \right)^2 \right],$$
$$\Rightarrow \qquad \dot u^2 = u^4 \left( \alpha (u) + a^2 u^2 \right) \Omega^2 \left[ \alpha \left( \gamma q + \beta l \right)^2 - 2 \beta \left( \gamma q + \beta l \right) \left( \alpha l - \beta q \right) - \gamma \left( \alpha l - \beta q \right)^2 \right].$$
Thus, we shall have:
\begin{equation}
\label{kerrtraj} \left( \frac{d u}{d \phi} \right)^2 = u^4 \left( \alpha (u) + a^2 u^2 \right) \left[ \alpha \left( \frac{\gamma q + \beta l}{\alpha l - \beta q} \right)^2 - 2 \beta \frac{\gamma q + \beta l}{\alpha l - \beta q} - \gamma \right].
\end{equation}
From which the following Binet's equation is derived.
\begin{equation}
\label{binetkerr} \frac{d^2 u}{d \phi^2} = \frac{d W (u)}{d u} , \quad \text{where } \quad W (u) = \left( \alpha (u) u^4 + a^2 u^6 \right) \left[ \alpha \left( \frac{\gamma q + \beta l}{\alpha l - \beta q} \right)^2 - 2 \beta \frac{\gamma q + \beta l}{\alpha l - \beta q} - \gamma \right].
\end{equation}
Under approximation upto 2nd order, we will have from (\ref{kerrtraj}):
$$\Rightarrow \qquad \frac{d \phi}{d u} \approx \frac1{l q} \left[ 1 - 2 GM \frac{aq}l u + \frac12 \left\{ 8 G^2 M^2 \left( 1 + \frac{a q}l + \left( \frac{aq}l \right)^2 \right) - a^2 \right\} u^2 \right]$$
Thus, for $K_1 = 2 G M u_0 \frac{aq}l, \ K_2 = \frac{(u_0)^2}2 \left\{ 8 G^2 M^2 \left( 1 + \frac{a q}l + \left( \frac{aq}l \right)^2 \right) - a^2 \right\}$ the deflection angle is given by:
$$\Delta \phi \approx \frac{u_0}{l q} \int_0^1 dv  \left[ 1 + (K_2 - K_1) v - K_2 v (1 - v) \right] = \frac{u_0}{l q} \left[ 1 + \frac{K_2 - K_1}2 - K_2 B (2 ,2) \right],$$
\begin{equation}
\label{kerrdefl} \mathcal D = 2 \Delta \phi - \pi = \frac{u_0}{l q} \left[ 1 + \frac{K_2 - K_1}2 - K_2 B (2 ,2) \right] - \pi.
\end{equation}
and the angle as a function of $u$ in terms of incomplete Beta functions is given by:
\begin{equation}
\label{kerrsol} \phi (u) = \frac1{lq} \int_0^u d x \left[ 1 - \frac{K_1}{u_0} x + \frac{K_2}{(u_0)^2} x^2 \right] = \frac1{lq} \left[ u - \frac{(K_1)^3 u_0}{(K_2)^2} B_y (2, 0) \right], \qquad u < u_0.
\end{equation}
where $y = - \frac{K_2}{K_1} \frac u{u_0}$.

\subsection{Null geodesics of the Tomimatsu-Sato 2 metric}

Here, we consider null-geodesics of the Tomimatsu-Sato solutions \cite{tomsato, tomsatop, bosewang, glass, tomsato1, tomsato2} and its various 
limits \cite{kinkel} in the equatorial plane, which are a series of exact solutions that include the Kerr 
metric, and other solutions. These solutions are characterised as being:

\begin{enumerate}
\item stationary,
\item axisymmetric,
\item asymptotically flat, and 
\item exact.
\end{enumerate}

\noindent
This new series of solutions contain three parameters: (a) mass $m$, (b) angular momentum $J \equiv m^2 q$, and (c) distortion parameter $\delta$. The distortion parameter $\delta$ specifies the Weyl solutions in static cases ($q = 0$). The Kerr 
and Tomimatsu-Sato soutions correspond to $\delta = 1$ and $\delta = 2$, respectively. \bigskip

\noindent
The Tomimatsu-Sato spacetime is given by:
\begin{equation}
\label{tomsato} \begin{split}
ds^2 &= - f \left( dt - \omega \ d \phi \right)^2 + f^{-1} \left[ \text{e}^{2 \gamma} \left( d \rho^2 + d z^2 \right) + \rho^2 \ d \phi^2 \right] \\
\text{where } \quad f &= \frac{A_\delta}{B_\delta}, \qquad \omega = 2mq \frac{C_\delta \left( 1 - y^2 \right)}{A_\delta}, \qquad \text{e}^{2 \gamma} = \frac{A_\delta}{p^{2 \delta} \left( x^2 - y^2 \right)^{\delta^2}}
\end{split}
\end{equation}
where $(\rho, \phi, z)$ describe cylindrical polar co-ordinates, and $x$ and $y$ are prolate spheroidal co-ordinates:
\begin{equation}
\label{prolsph} \rho = \frac{m p}{\delta} \sqrt{\left( x^2 - 1 \right) \left( 1 - y^2 \right)}, \qquad z = \frac{m p}{\delta} x y.
\end{equation}
and $p$ and $q$ are rotation parameters. Here, we shall deal with the case $\delta = 2$. As we can 
see from (\ref{prolsph}), for $\rho$ to be real, $y^2 \leq 1 \leq x^2$, meaning that $-1 \leq y \leq 1$. So to get $z = 0$ for geodesics in the equatorial plane, we are confined to only one choice:
\begin{equation} \label{eqplset}
z = 0 \qquad \Rightarrow \qquad y = 0 \quad \Rightarrow \quad \rho = \dfrac{mp}2 \sqrt{x^2 - 1} \qquad \rho \in \mathbb R.
\end{equation}
\begin{figure}
\begin{center}
\includegraphics[width=0.25\linewidth]{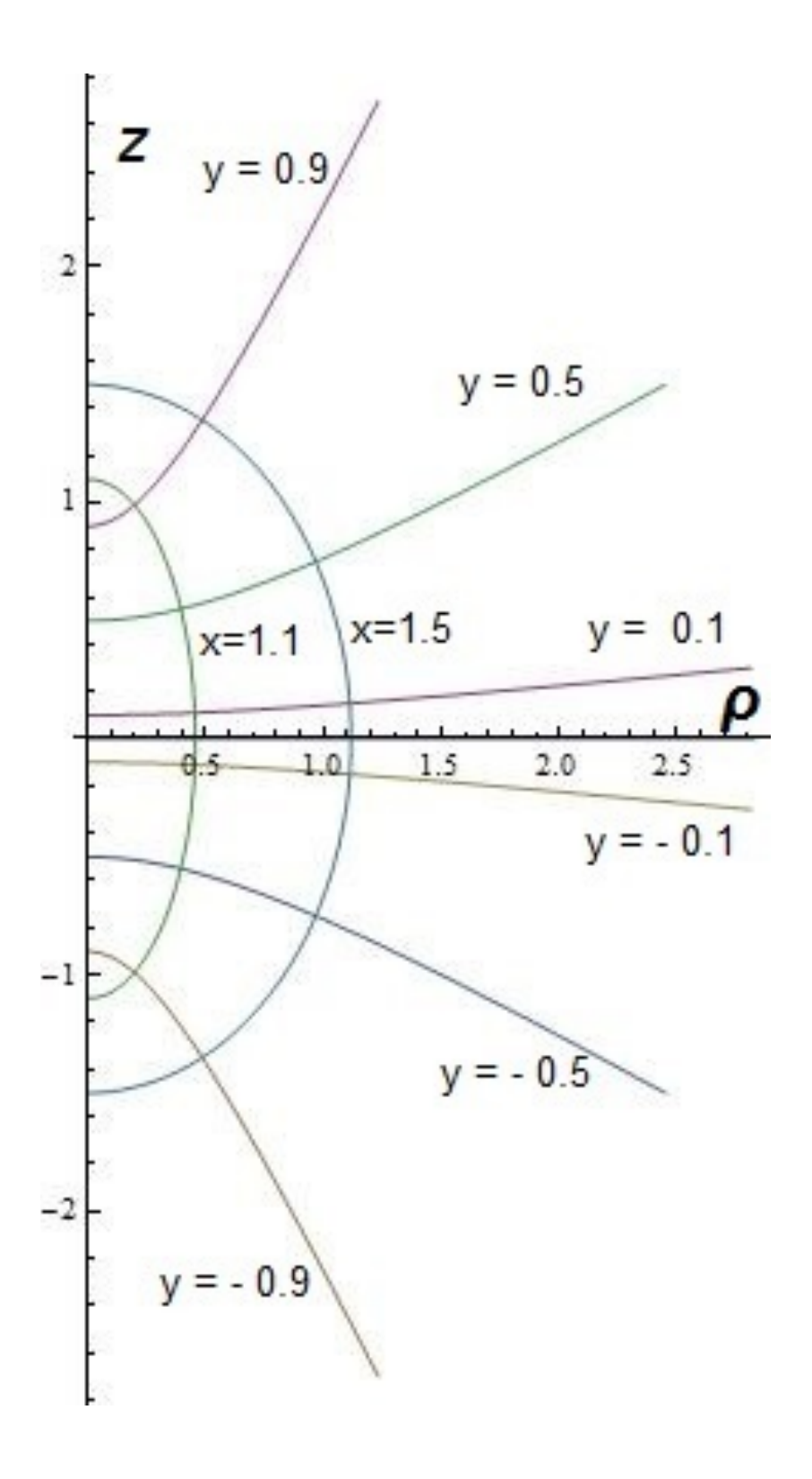}
\caption{\small Various curves of the Tomimatsu-Sato metric for constant $x = 1.1, 1.5$, and constant $y = \pm 0.5, \pm 0.9$}
\end{center}
\end{figure}
It is not possible to choose $x = 0$ if we are to consider real values of $\rho$, because then $x$ has to cross the boundary $x^2 = 1$, at which point, the radius becomes imaginary. Choosing $y = 0$, according to (\ref{eqplset}) lets us write the functions from (\ref{tomsato}) in the equatorial plane for $n = 2$ as: 
\begin{equation}
\label{eqfunc1} 
x = \frac{\sqrt{4 \rho^2 + m^2 p^2}}{mp} \qquad \Rightarrow \qquad f = \frac{A_2}{B_2}, \qquad \omega = 2mq \frac{C_2}{A_2}, \qquad \text{e}^{2 \gamma} = \frac{A_2}{p^4 x^8}
\end{equation}
where the functions $A, B, C$ for $\delta = 2$ are given from \cite{tomsatop} by:
\begin{equation} \label{abcf}
\begin{split}
A_2 &= \left[ p^2 (x^2 - 1)^2 + q^2 (1 - y^2)^2 \right]^2 - 4 p^2 q^2 (x^2 - 1) (1 - y^2) (x^2 - y^2)^2 \\
B_2 &= \left[ p^2 (x^4 - y^4) - (1 - y^4) + 2 p x (x^2 - 1) \right]^2 + 4 q^2 y^2 \left[ p x (x^2 - y^2) + 1 - y^2 \right]^2 \\
C_2 &= - 2 p^2 x (x^2 - 1) (x^2 - y^2) (p x^2 + 2 x + p) + (1 + px) (1 - y^2) \left[ p^2 (x^2 - y^2) (x^2 + y^2 - 2) + (1 - y^2)^2 \right],
\end{split}
\end{equation}
where for $y = 0$, we should have the functions listed in \cite{bosewang} from the functions (\ref{abcf}) defined 
in \cite{glass, hikod}:
\begin{equation} \label{abc} y = 0 \quad \Rightarrow \quad 
\begin{cases}
A_2 &= [ p^2 (x^2 - 1)^2 + q^2]^2 - 4 p^2 q^2 (x^2 - 1) x^4 \\
&= p^4 x^8 \left[ \left\{ p^{-2} x^{-4} + (1 - 2 x^{-2}) \right\}^2 + 4 x^{-2} (1 - p^{-2}) (1 - x^{-2} ) \right] \\
B_2 &= (p^2 x^4 + 2 p x^3 - 2 p x - 1)^2 \\
&= p^4 x^8 \left( 1 + 2 p^{-1} x^{-1} - 2 p^{-1} x^{-3} - p^{-2} x^{-4} \right)^2 \\
C_2 &= - 2 p^2 x^2 (x^2 - 1)^2 - p^2 x^4 (1 + px) (2 x^2 - 1) + (1 + px) \\
&= p^4 x^8 \left[ - 2 p^{-2} x^{-2} (1 - x^{-2} )^2 - p^{-1} x^{-1} (1 + p^{-1} x^{-1}) (2 - x^{-2}) + p^{-3} x^{-7} (1 + p^{-1} x^{-1}) \right].
\end{cases}
\end{equation}
where $p^2 + q^2 = 1$, and $\omega_0$ is independent of $\delta$. Using (\ref{abc}) in (\ref{eqfunc1}), after isolating a common factor 
$p^4 x^8$ from all terms, we will get on taking approximations upto the 2nd order:
$$x^{-1} = \left[ \left( \frac{2 \rho}{m p} \right)^2 + 1 \right]^{-\frac12} \approx \frac{m p}{2 \rho} \qquad \Rightarrow \qquad 2 p^{-1} x^{-1} \approx m u,$$
\[ \begin{split}
f = \frac{A}{B} &= \frac{\left[ p^{-2} x^{-4} + (1 - 2 x^{-2}) \right]^2 + 4 x^{-2} (1 - p^{-2}) (1 - x^{-2} )}{\left( 1 + 2 p^{-1} x^{-1} - 2 p^{-1} x^{-3} - p^{-2} x^{-4} \right)^2} \\
& \approx \frac{1 - 4 x^{-2} p^{-2}}{\left( 1 + 2 p^{-1} x^{-1} \right)^2} = \frac{1 - 2 x^{-1} p^{-1}}{1 + 2 p^{-1} x^{-1}} \approx \frac{\rho - m}{\rho + m}
\end{split} \]

\[ \begin{split}
\omega = 2mq \frac{C}{A} &= 2 m q \frac{- 2 p^{-2} x^{-2} (1 - x^{-2} )^2 - p^{-1} x^{-1} (1 + p^{-1} x^{-1}) (2 - x^{-2}) + p^{-3} x^{-7} (1 + p^{-1} x^{-1})}{\left[ p^{-2} x^{-4} + (1 - 2 x^{-2}) \right]^2 + 4 x^{-2} (1 - p^{-2}) (1 - x^{-2} )} \\
&\approx - 2 m q \frac{2 p^{-1} x^{-1} (1 + 2 p^{-1} x^{-1})}{1 - 4 x^{-2} p^{-2}} = - 2 m q \frac{2 p^{-1} x^{-1}}{1 - 2 p^{-1} x^{-1}} \approx - 2 m q \frac m{\rho - m}
\end{split} \]

\[ \begin{split}
\text{e}^{2 \gamma} = \frac{A}{p^4 x^8} &= \left[ p^{-2} x^{-4} + (1 - 2 x^{-2}) \right]^2 + 4 x^{-2} (1 - p^{-2}) (1 - x^{-2} ) \\
&\approx 1 - 4 p^{-2} x^{-2} \approx 1 - \left( \frac m{\rho} \right)^2,
\end{split} \]
\begin{equation}
\label{eqfunc2} \therefore \qquad f = \frac{\rho - m}{\rho + m}, \qquad \omega = - \frac{2m^2 q}{\rho - m}, \qquad \text{e}^{2 \gamma} = 1 - \frac{m^2}{\rho^2}.
\end{equation}
The conserved quantities deduced from the metric (\ref{tomsato}) are given by:
\begin{equation}
\label{cons} \begin{split}
\varepsilon &= - f \dot t + f \omega \dot \phi, \\ 
l &= f \omega \dot t + f \left( \frac{\rho^2}{f^2} - \omega^2 \right) \dot \phi
\end{split} \qquad \Rightarrow \qquad 
\begin{split}
\dot t &= f \frac{\omega \left( l + \omega \varepsilon \right) - \frac{\rho^2}{f^2} \varepsilon}{\rho^2} \\
\dot \phi &= f \frac{l + \omega \varepsilon}{\rho^2}
\end{split}
\end{equation}
For a null-geodesic in a plane with $\dot z = 0$, we have:
$$\left( \dot t - \omega \dot \phi \right)^2 = f^{-2} \left[ \text{e}^{2 \gamma} \dot \rho^2 + \rho^2 \dot \phi^2 \right] \qquad \Rightarrow \qquad \frac{\varepsilon^2}{f^2} = f^{-2} \text{e}^{2 \gamma} \dot \rho^2 + \frac{(l + \omega \varepsilon)^2}{\rho^2}.$$
Writing $b = \frac l{\varepsilon}$, we will have:
$$\dot \rho^2 = \varepsilon^2 \text{e}^{- 2 \gamma} \left[ 1 - f^2 \frac{(b + \omega)^2}{\rho^2} \right], \qquad \qquad \dot \phi = \varepsilon f \frac{(b + \omega)}{\rho^2},$$
\begin{equation}
\label{tomsatnull} \Rightarrow \qquad \left( \frac{d u}{d \phi} \right)^2 = \text{e}^{- 2 \gamma} \left[ \frac1{f^2 (b + \omega)^2} - u^2 \right] \qquad \text{where} \quad u = \frac1{\rho}.
\end{equation}
For circular null-geodesics, we would have to have:
\begin{equation}
\label{circsol} \frac{d \rho}{d \phi} = 0 \quad \Rightarrow \quad f = \frac{\rho}{b + \omega}.
\end{equation}
Thus, the Binet's equation for Tomimatsu-Sato metric is:
\begin{equation}
\label{tsbinet} \frac{d^2 u}{d \phi^2} = \text{e}^{- 2 \gamma} \left[ \frac{d \ }{d u} \left( \frac1{f^2 (b + \omega)^2} \right) - \frac{d \gamma}{d u} \left( \frac1{f^2 (b + \omega)^2} - u^2 \right) - u \right].
\end{equation}
The solutions to Binet's equation for Tomimatsu-Sato metric (\ref{tsbinet}) are given by:
\begin{equation}
\label{solts} \Delta \phi = \int_0^{u_0} \left[ \frac1{f^2 (b + \omega)^2} - u^2 \right]^{-\frac12} \text{e}^{\gamma} du.
\end{equation}
where $u_0$ is the maximum value of $u$ where $\left( \frac{d u}{d \phi} \right)_{u_0} = 0$. We shall now consider the various limits applicable to this metric.

\subsubsection{Weak field limit}

There are various limits that can be applied to the Tomimatsu-Sato metric to derive special solutions. In 
the weak-field limit where $m \rightarrow 0$ \cite{kinkel}, the angular momentum $J = m^2 q$ could vanish. In this case, the 
extended Tomimatsu-Sato solution is needed, requiring that we define a Kerr parameter $a$ in $p$ and $q$ as:
\begin{equation}
\label{kerrpar} q = \frac am, \quad \Rightarrow \quad p = \sqrt{1 - q^2} = \frac{\sqrt{m^2 - a^2}}m \qquad (\because \ p^2 + q^2 = 1)
\end{equation}
Thus, the deflection angle solution for weak field approximation will be the same as (\ref{tomsatdef}) 
taking approximation upto 2nd order. If we apply (\ref{kerrpar}) into (\ref{eqfunc2}), we will have:
\begin{equation}
\label{eqfunc3} f \approx 1 - 2m u + 2m^2 u^2, \qquad \omega \approx - 2a m u \left( 1 + mu \right), \qquad \text{e}^{2 \gamma} \approx 1 - m^2 u^2,
\end{equation}
$$\frac{d \phi}{d u} \approx b - 2 m (a + b) u + \frac{m^2}2 \left( 3 b + 4 a + \frac{b^3}{m^2} \right) u^2.$$
If we write $K_1 = 2 m u_0 (m + b), K_2 = \frac{m^2 u_0^2 b}2 \left( 3 b + 4 a + \frac{b^3}{m^2} \right)$, then we have:
\[ \begin{split}
\Delta \phi &= \int_0^1 \left[ b u_0 - K_1 v + K_2 v^2 \right] dv, \qquad \qquad \text{where } \  v = \frac u{u_0} \\ 
&= b u_0 + (K_2 - K_1) \int_0^1 dv - K_2 \int_0^1 dv \ v ( 1 - v ) = b u_0 + \frac{K_2 - K_1}2 - K_2 B (2, 2). \\
\end{split} \] 
Thus, the deflection angle for null-geodesics here is given by:
\begin{equation}
\label{tomsatdef} \mathcal D = 2 \Delta \phi - \pi \ = \ 2 b u_0 + \frac{K_2 - K_1}2 - K_2 B (2, 2) - \pi.
\end{equation}
The solution using Chebyshev's theorem is:
\begin{equation}
\label{chebtomsat} \theta (u) = \int_0^u dx \left[ b - \frac{K_1}{u_0} x + \frac{K_2}{u_0^2} x^2 \right] = b u - \frac{(K_1)^3 u_0}{(K_2)^2} B_y ( 2, 0 ), \qquad u < u_0
\end{equation}
where $y = - \frac{K_2}{K_1} \frac u{u_0}$. This concludes the solution of the null-geodesic using incomplete Beta functions.

\subsubsection{Disc model}

Now there is a way to allow more freedom to the value of $x$, while keeping $\rho$ real. Normally, if we allowed 
$x \leq 1$, then $\rho$ would have imaginary values. However, if we decide to permit $p$ to have imaginary values, 
then we shall see on allowing $x^2 \leq 1$ in (\ref{prolsph}):
$$p = - i \hat p, \hat p \in \mathbb R, \ x^2 \leq 1 \quad \Rightarrow \quad \rho = \frac{m (ip)}{\delta} \sqrt{ \left( 1 - x^2 \right) \left( 1 - y^2 \right)} = \frac{m \hat p}{\delta} \sqrt{ \left( 1 - x^2 \right) \left( 1 - y^2 \right)} \in \mathbb R.$$
However, this would result in $z$ being imaginary:
$$z = \frac{m p}{\delta} x y = - i \frac{m \hat p}{\delta} x y \ \in \ \mathbb I.$$
Thus, we must go one step further, and demand that $x$ be imaginary itself for $z$ to be real, while also 
keeping $\rho$ real. Upon applying this step to (\ref{prolsph}), we finally get:
\begin{equation} \label{adjcord}
\begin{split} p &= - i \hat p, x = i \hat x \\ 
&\hat x, \hat p \in \mathbb R \end{split} \quad \Rightarrow \quad 
\begin{cases}
\rho &= \frac{m p}{\delta} \sqrt{- \left( \hat x^2 + 1 \right) \left( 1 - y^2 \right)} = \frac{m \hat p}{\delta} \sqrt{\left( \hat x^2 + 1 \right) \left( 1 - y^2 \right)} \\
z &= \frac{m p}{\delta} x y = \frac{m \hat p}{\delta} \hat x y
\end{cases} \quad \in \mathbb R
\end{equation}
Finally, we have full freedom for the value of $\hat x$, allowing us to even reach $\hat x = 0$, while keeping both $\rho$ and $z$ real. The only restriction is that both $x$ and $p$ be imaginary, as shown in \cite{kinkel}. 
\begin{figure}
\begin{center}
\includegraphics[width=0.34\linewidth]{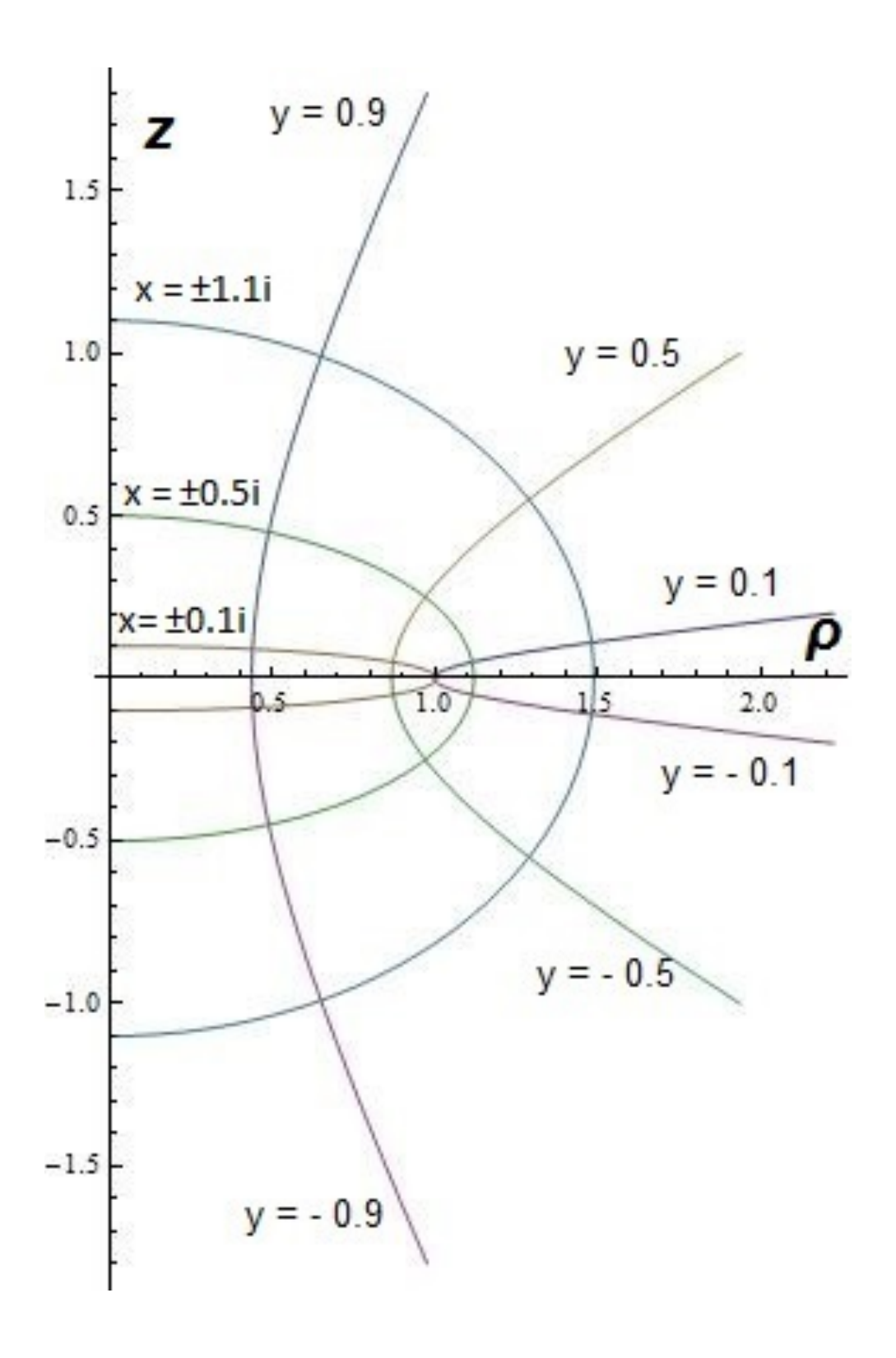}
\caption{\small Curves of the TS metric for disc model with constant $\hat x = 0.1, 0.5, 1.1$, and $y = \pm 0.1. \pm 0.5, \pm 0.9$}
\label{fig:disc}
\end{center}
\end{figure}

Such freedom cannot be afforded to $y$ since writing $p = - i \hat p, y = i \hat y$ leaves $\rho$ imaginary. Even if we keep 
$x, y \in \mathbb R$, and $p = - i \hat p$, then $\rho$ could be real for $y^2 \geq 1$, but $z$ becomes imaginary and a new restriction is 
imposed on the range of values of $y$. Thus, the setting $p = - i \hat p, x = i \hat x$ is the only way to allow freedom 
of values to $x$. Naturally, when $x = 0$, due to the limitation $y^2 \leq 1$, we will have the metric defined on a 
disc according to (\ref{adjcord})
\begin{equation}
\label{disc} \hat x = 0, \ y^2 \leq 1 \qquad \Rightarrow \qquad z = 0, \quad \rho = \frac{m \hat p}{\delta} \sqrt{1 - y^2} \leq \frac{m \hat p}{\delta} \in \mathbb R,
\end{equation}
while having $x = 0$ defines the metric outside the disc
\begin{equation}
\label{hole} y = 0 \qquad \Rightarrow \qquad z = 0, \quad \rho = \frac{m \hat p}{\delta} \sqrt{\hat x^2 + 1} \geq \frac{m \hat p}{\delta} \in \mathbb R.
\end{equation}
Naturally, in this model, setting $\hat x = y = 0$ in (\ref{adjcord}), will describe a ring:
$$\rho = \frac{m \hat p}{\delta} \sqrt{\left( 0 + 1 \right) \left( 1 - 0 \right)} = \frac{m \hat p}{\delta}.$$
If we consider the functions defined in \cite{tomsatop} for $\delta = 2$, then for $x = i \hat x, \ p = - i \hat p$, we will have:
\begin{equation} \label{discf1}
\begin{split}
A &= \left[ (1 - y^2 )^2 - \hat p^2 ( \hat x^2 + y^2 ) (2 + \hat x^2 - y^2) \right]^2 - 4 \hat p^2 (1 + \hat p^2) (1 + \hat x^2 ) (1 - y^2) (\hat x^2 + y^2)^2 \\
B &= \left[ \hat p^2 ( \hat x^4 - y^4 ) + (1 - y^4) + 2 \hat p \hat x (1 + \hat x^2) \right]^2 + 4 (1 + \hat p^2) y^2 \left[ (1 - y^2) - \hat p \hat x ( \hat x^2 + y^2 ) \right]^2 \\
C &= - 2 \hat p \hat x (1 + \hat x^2) (\hat x^2 + y^2) \left[2 \hat p \hat x - \hat p^2 (1 - \hat x^2) \right] \\
& \quad + \hat p^2 (1 + \hat p \hat x) (1 - y^2) (\hat x^2 + y^2) (y^2 - \hat x^2 - 2) + (1 + \hat p \hat x) (1 - y^2)^3.
\end{split}
\end{equation}
Now we shall consider the two ways to define null-geodesics in the equatorial plane under this new setting.

\subsubsection*{$\bm{y = 0}$ region}

Now, from \cite{kinkel}, we can say that for $y = 0$, we have:
\begin{equation}
\label{ernstpot} \begin{split}
\xi_2 &= \frac{- 2 \hat p \hat x (\hat x^2 + 1) - 2 i q y (1 - y^2)}{- \hat p^2 (\hat x^4 - 1) + 2 i \hat p q \hat x y (\hat x^2 + y^2) - q^2 (1 - y^4)}, \qquad \xi_{2 (y = 0)} = \frac{2 \hat p \hat x ( 1 + \hat x^2 )}{\hat p^2 \hat x^4 + 1} \approx \frac m{\rho} = m u, \\ 
&\Rightarrow \qquad f = \text{Re} \left( \frac{1 - \xi}{1 + \xi} \right) \approx (1 - m u) (1 - mu + m^2 u^2) \approx 1 - 2 m u + 2 m^2 u^2.
\end{split}
\end{equation}
Setting $y = 0$ in (\ref{discf1}), and using (\ref{hole}) for $\delta = 2$, we will have the following functions:
\[ \begin{split}
A &= \left[ 1 - \hat p^2 \hat x^2 (2 + \hat x^2) \right]^2 - 4 \hat p^2 \hat x^4 (1 + \hat p^2) (1 + \hat x^2 ) \\
B &= \left[ \hat p^2 \hat x^4 + 1 + 2 \hat p \hat x (1 + \hat x^2) \right]^2 \\
C &= - 2 \hat p \hat x^3 (1 + \hat x^2) \left[2 \hat p \hat x - \hat p^2 (1 - \hat x^2) \right] + (1 + \hat p \hat x) \left[ 1 - \hat p^2 \hat x^2 (\hat x^2 + 2) \right] .
\end{split} \]
Writing $u = \rho^{-1}$, we have taking terms only upto 2nd order since $m u << 1$,
$$\hat x^{-1} = \left[ \left( \frac{2 \rho}{m \hat p} \right)^2 - 1 \right]^{-\frac12} \approx \frac{m \hat p}{2 \rho} \qquad \Rightarrow \qquad 2 \hat p^{-1} \hat x^{-1} \approx m u,$$
\[ \begin{split}
f = \frac{A}{B} &= \frac{\left[ \hat p^{-2} \hat x^{-4} - (1 + 2 \hat x^{-2}) \right]^2 - 4 \hat x^{-2} (1 + \hat p^{-2}) (1 + \hat x^{-2} )}{\left[ 1 + \hat p^{-2} \hat x^{-4} + 2 \hat p^{-1} \hat x^{-1} (1 + \hat x^{-2}) \right]^2} \\
& \approx 1 - 4 \hat p^{-1} \hat x^{-1} + 8 \hat p^{-2} \hat x^{-2} \approx 1 - 4 \hat p^{-1} w + 8 \hat p^{-2} w^2 = 1 - 2 m u + 2 (m u)^2
\end{split} \]

\[ \begin{split}
\omega = 2mq \frac{C}{A} &= 2 m q \frac{- 2 \hat p^{-1} \hat x^{-1} (1 + \hat x^{-2}) \left( 1 + 2 \hat p^{-1} \hat x^{-1} - \hat x^{-2} \right) - \hat p^{-1} \hat x^{-3} (1 + \hat p^{-1} \hat x^{-1}) (1 + 2 \hat x^{-2} - \hat p^{-2} \hat x^{-4})}{\left[ \hat p^{-2} \hat x^{-4} - (1 + 2 \hat x^{-2}) \right]^2 - 4 \hat p^{-2} \hat x^{-2} (1 + \hat p^2) (1 + \hat x^{-2} )} \\
&\approx - 4 m q \hat p^{-1} \hat x^{-1} \left( 1 + 2  \hat p^{-1} \hat x^{-1} \right) \approx - 4 m q \hat p^{-1} w \left( 1 + 2 \hat p^{-1} w \right) \\
&= - 2 m^2 q u \left( 1 + m u \right)
\end{split} \]

\[ \begin{split}
\text{e}^{2 \gamma} = \frac{A}{\hat p^4 x^8} &= \left[ \hat p^{-2} \hat x^{-4} - (1 + 2 \hat x^{-2}) \right]^2 - 4 \hat p^{-2} \hat x^{-2} (1 + \hat p^2) (1 + \hat x^{-2} ) \\
&\approx 1 - 4 \hat p^{-2} \hat x^{-2} \approx 1 - 4 \hat p^{-2} w^2 = 1 - \left( m u \right)^2,
\end{split} \]
which are essentially the same as those for the weak field limit provided by (\ref{eqfunc3}). Thus, the deflection 
angle and null-geodesic solution will be the same as in the case of the weak-field limit for $y = 0$. On 
the other hand, for $\hat x = 0$, we will have $\rho \leq \frac{m \hat p}2$, meaning that the weak field approximation is not 
applicable.

\subsubsection*{$\bm{\hat x = 0}$ region}

We can see from the fig \ref{fig:disc} that for $\hat x = 0$, we will have a discontinuity in the value of $y$ as we cross the 
disc plane defined by $\hat x = 0$. Thus, it would be preferable to apply a limit that brings us as close to the 
disc as possible. \bigskip

\noindent
For $\hat x = \varepsilon, \ y = 0$, and $\hat p = finite$, we shall have from (\ref{discf1}):
\[ \begin{split}
A &= 1 - 2 \hat p^2 \varepsilon^2 (2 + \varepsilon^2) + \hat p^4 \varepsilon^4 (2 + \varepsilon^2)^2 - 4 \hat p^2 (1 + \hat p^2) \varepsilon^4 (1 + \hat p^2) (1 + \varepsilon^2 ) \xrightarrow{\varepsilon \rightarrow 0} 1 \\
B &= \left[ \hat p^2 \varepsilon^4 + 1 + 2 \hat p \varepsilon (1 + \varepsilon^2) \right]^2 \xrightarrow{\varepsilon \rightarrow 0} 1 \\
C &= - 2 \hat p \varepsilon^3 (1 + \varepsilon^2) \left[2 \hat p \varepsilon - \hat p^2 (1 - \varepsilon^2) \right] - \hat p^2 \varepsilon^2 (1 + \hat p \varepsilon) (\varepsilon^2 + 2) + (1 + \hat p \varepsilon) \xrightarrow{\varepsilon \rightarrow 0} 1.
\end{split} \]
Thus, since none of the functions vanish for $y = 0, \ \varepsilon \rightarrow 0$, we can ignore all $\varepsilon^2$ terms, leading us to the following approximate values of the functions on the disc using (\ref{disc})
\begin{equation} \label{discf2}
\begin{split}
A &= \left[ (1 - y^2 )^2 - \hat p^2 y^2 (2 - y^2) \right]^2 - 4 \hat p^2 y^2 (1 + \hat p^2) (1 - y^2) \\ 
&= \left[ \left( \frac{2 \rho}{m \hat p} \right)^4 (1 + \hat p^2) - \hat p^2 \right]^2 - 4 \hat p^2 (1 + \hat p^2) \left( \frac{2 \rho}{m \hat p} \right)^2 \left[ 1 - \left( \frac{2 \rho}{m \hat p} \right)^2 \right] \\
B &= \left[ (1 + \hat p^2) y^4 - 1 \right]^2 + 4 (1 + \hat p^2) y^2 (1 - y^2) \\
&= \left[ (1 + \hat p^2) \left\{ 1 - \left( \frac{2 \rho}{m \hat p} \right)^2 \right\}^2 - 1 \right]^2 + 4 (1 + \hat p^2) \left( \frac{2 \rho}{m \hat p} \right)^2 \left[ 1 - \left( \frac{2 \rho}{m \hat p} \right)^2 \right] \\
C &= \hat p^2 y^2 (1 - y^2) (y^2 - 2) + (1 - y^2)^3 \\
&= - \hat p^2 \left( \frac{2 \rho}{m \hat p} \right)^2 \left\{ 1 - \left( \frac{2 \rho}{m \hat p} \right)^4 \right\} + \left( \frac{2 \rho}{m \hat p} \right)^6.
\end{split}
\end{equation}
and the metric functions can be deduced by applying (\ref{discf2}) to (\ref{tomsato}), thus allowing us to deduce 
null-geodesics within the disc. For positions closer towards the center, $\rho \rightarrow 0$, defined by $y \rightarrow 1$, we shall see that according to (\ref{tomsato}), we will have:
\[ \begin{split}
\lim_{y \rightarrow 1} A &= \hat p^4, \\
\lim_{y \rightarrow 1} B &= \hat p^4, \\
\lim_{y \rightarrow 1} C &= 0, 
\end{split} \qquad \Rightarrow \qquad 
\begin{split}
\lim_{y \rightarrow 1} f &= \lim_{y \rightarrow 1} \frac{A}{B} = 1, \\
\lim_{y \rightarrow 1} \omega &= \lim_{y \rightarrow 1} 2mq (1 - y^2) \frac{C}{A} = 0, \\
\lim_{y \rightarrow 1} \text{e}^{2 \gamma} &= \lim_{y \rightarrow 1} \frac{A}{p^4 y^8} = 1,
\end{split} \]
thus, showing that closer towards the center, the metric is nearly flat.

\subsubsection{q = 1 limit}

For the setting $p = 0, q = 1$, we are essentially considering the extreme Kerr limit $a = m$ which 
coincides with the Tomimatsu-Sato metric . However, the metric deduced from this setting is 
not the extreme Kerr due to the singularity that arises in the transformation $(\rho, z) \rightarrow (x, y)$ 
when $p \rightarrow 0$ in (\ref{prolsph}). This is evident if we try to set $p = 0$ in (\ref{eqfunc1}). \smallskip

\noindent
One way around this, as suggested in \cite{kinkel} is to demand that 
\begin{equation}
\label{demand} px^\alpha \sim \left( \frac{\rho}m \right)^\alpha = finite, \qquad \alpha = 2k - 1.
\end{equation}
This is easily demonstrated by the example of taking (\ref{eqfunc1}), and demanding that for $\alpha = 1$, $px = finite$. ie.
$$\lim_{p \rightarrow 0} px = \frac1m \lim_{p \rightarrow 0} \sqrt{4 \rho^2 + m^2 p^2} = 2 \frac{\rho}m.$$
implying that in this limit, the permitted radius $\rho$ for geodesics is immensely larger than $mp/2$. 
The appropriate co-ordinate replacement is given by:
$$\left( \frac{\rho}m \right)^{2k-1} = \frac{(k!)^2}{2k!} p x^{2k-1}, \qquad y = \cos \theta.$$
From \cite{kinkel}, we will have the functions in the equatorial plane for $y = 0$ and $k = 1$ given by:
\begin{equation}
\label{eqfunc5} f = 2 \frac AB, \qquad \omega = \rho \frac CA, \qquad \text{e}^{2 \gamma} = A \left( \frac{\rho}m \right)^{-2},
\end{equation}
where according to \cite{kinkel}, for $y = 0$ and $k = 1$, we get 
\begin{equation} \label{eqfunc6}
A = \left( \frac{\rho}m \right)^2 - 1, \qquad B = 2 \left( \frac{\rho}m - 1 \right)^2, \qquad C = 2 \left( \frac{\rho}m - 1 \right)
\end{equation}
\[ \begin{split}
\therefore
\end{split} \qquad 
\begin{split}
f &= \frac{\rho + m}{\rho - m} \approx 1 + 2 \frac m{\rho} + 2 \left( \frac m{\rho} \right)^2, \\ 
\omega &= m \frac{\rho}{\rho + m} \approx m \left[ 1 - \frac m{\rho} + \left( \frac m{\rho} \right)^2 \right],
\end{split} \qquad \qquad 
\begin{split}
\text{e}^{2 \gamma} = 1 - \left( \frac m{\rho} \right)^2.
\end{split} \]
Applying these functions to (\ref{tomsatnull}), we have for aprroximation upto 2nd order in $u = \frac1{\rho}$:
$$\left( \frac{d u}{d \phi} \right)^2 = \left[ 1 - ( mu )^2 \right]^{-1} \left[ \frac1{f^2 (b + \omega)^2} - u^2 \right] \qquad \text{where} \quad u = \frac1{\rho}.$$
$$\Rightarrow \qquad \frac{d \phi}{d u} \approx (b + m) \left[ 1 + \frac{2b + m}{2(b + m)} m u - \left\{ \frac{5m}{8 (b + m)} - \frac{b^2 (b^2 + 4bm - 2m^2)}{2 m^2 (b + m)^2} \right\} (m u)^2 \right].$$
Thus, writing $\mathcal C_1 = \frac{2b + m}{2(b + m)} m u_0, \ \mathcal C_2 = \left\{ \frac{5m}{8 (b + m)} - \frac{b^2 (b^2 + 4bm - 2m^2)}{2 m^2 (b + m)^2} \right\} (u_0)^2$ the overall deflection angle is given in terms of beta functions as:
\[ \begin{split}
\Delta \phi &= (b + m) u_0 \int_0^1 d v \left(1 + \mathcal C_1 v - \mathcal C_2 v^2 \right), \qquad \qquad \text{where } \ v = \frac u{u_0} \\ 
&= (b + m) u_0 \left[ 1 + \left( \mathcal C_1 - \mathcal C_2 \right) \int_0^1 dv \ v + \mathcal C_2 \int_0^1 d v \ v \left(1 - v \right) \right] \\ 
&= (b + m) u_0 \left[ 1 + \frac{\mathcal C_1 - \mathcal C_2}2 + \mathcal C_2 B (2, 2) \right]
\end{split} \]
\begin{equation}
\label{tomsatlimdef} \mathcal D = 2 \Delta \phi - \pi = 2 (b + m) u_0 \left[ 1 + \frac{\mathcal C_1 - \mathcal C_2}2 + \mathcal C_2 B (2, 2) \right] - \pi.
\end{equation}
On the other hand, $\phi (u)$ as a function of $u$ in terms of incomplete beta functions is:
\begin{equation}
\label{tomsatlimsol} \phi (u) = (b + m) \int^u dx \left[ 1 + x \left( \frac{\mathcal C_1}{u_0} - \frac{\mathcal C_2}{(u_0)^2} x \right) \right] = (b + m) \left[ u + \frac{(\mathcal C_1)^3 u_0}{ ( \mathcal C_2 )^2} B_y ( 2, 0 ) \right], \qquad \quad y = - \frac{\mathcal C_2}{\mathcal C_1} \frac x{u_0}.
\end{equation}
which concludes the analysis of null-geodesics of Tomimatsu-Sato metrics in the $q = 1$ limit.

\numberwithin{equation}{section}

\section{Duality under conformal transformation}

A conformal transformation that preserves the Jacobi metric will reveal potential power law dualities. If we use complex variables to describe the planar co-ordinates as $z = x + iy$, then we can describe the Jacobi metric as:
\begin{equation}
\label{jac} ds^2_J = 2m \left( E - V(|z|) \right) d \bar{z} \ dz.
\end{equation}
If we employ the pullback with the conformal map $z \longrightarrow w = w (z)$, we will get the projective dual \cite{mcar} of (\ref{jac}):
\begin{equation} \label{map}
ds^2_J \xrightarrow{z \longrightarrow w = w (z)} 2m \left( \widetilde{E} - \widetilde{V} (|w|) \right) d \bar{w} \ dw \quad \Rightarrow \quad  
\begin{cases}
\widetilde{E} = V (|z|) \left| w' (z) \right|^{-2} \smallskip \\ 
\widetilde{V} (|w|) = E \left| w' (z) \right|^{-2}
\end{cases}.
\end{equation}
Let us consider only conformal maps of the form $w = z^p$. This means that according to (\ref{map}) the potential has to be :
$$V \propto |z|^a \qquad \qquad \big| w' (z) \big| \propto z^{p-1} \qquad \qquad \Rightarrow \qquad a = 2 (p - 1),$$
$$V (|z|) \propto |z|^{2p - 2}.$$
Conversely, this means that 
$$\widetilde{V} (|w|) \propto |z|^{2-2p} = |w|^{\frac{2-2p}p}.$$
where, for various settings of $p$, we will get various dualities that preserve the form of the non-relativistic Jacobi metric:
\begin{enumerate}

\item For $p = 2$, we essentially get the Kepler-Hooke duality. 
$$V (|z|) \propto |z|^2 \qquad \qquad \widetilde{V} (|w|) = |w|^{-1}.$$
also known as the Bohlin-Arnold duality, an equivalence between the Kepler and Hooke 
mechanical systems in the plane originating in a paper by Bohlin \cite{bohlin}, and Arnold \cite{arnold}. Let 
us apply this co-ordinate conversion as $w: (r, \phi) \longrightarrow (y = r^2, \varphi = 2 \phi)$. This should let 
us write from the null version of Lagrangian (\ref{lag}):
$$- \frac{f(r) \dot{t}^2}{\dot{\phi}^2} + \frac1{g(r)} \left( \frac{\dot{r}}{\dot{\phi}} \right)^2 + r^2 = 0 \qquad \Rightarrow \qquad - g(r) \frac{f(r) \left( r^2 \right)^3 \dot{t}^2}{\left( r^2 \dot{\phi} \right)^2} + \left( r \frac{d r}{d \phi} \right)^2 + g(r) \left( r^2 \right)^2 = 0,$$
$$\Rightarrow \qquad \left( \frac{d y}{d \varphi} \right)^2 + y^2 = y^3 \frac{\widetilde{g}(y)}{\widetilde{f}(y)} \frac{q^2}{l^2} - y^2 \widetilde{G}(y).$$
where for $f(r) = g(r)$ and $G(r) \propto r^{2-n} = y^{1 - \frac n2}$ and $A = 2 M_n$, $B = b^{-2}$ for the 
Schwarzschild-Tangherlini metric (\ref{schtan}), the equation in original co-ordinates is:
\begin{equation}
\label{orgnl} \left( \frac{d r}{d \phi} \right)^2 + r^2 = A r^{4-n} + B r^4.
\end{equation}
and, under conformal transformation $w: (r, \phi) \longrightarrow (y = r^2, \varphi = 2 \phi)$, we get:
$$\left( \frac{d y}{d \varphi} \right)^2 + y^2 = A y^3 + B y^{3 - \frac n2}.$$
Upon setting $n=6$ in the above result, we will see that (\ref{orgnl}) is dual to 
$$\left( \frac{d y}{d \varphi} \right)^2 + y^2 = A y^3 + B.$$
which is the equation (\ref{mde3}) with $n=3$.

\item For $p = -1$, we get self-duality. 
$$\widetilde{V} (|w|) \sim V (|z|) \propto |z|^{-4} = |w|^{-4}.$$
Now, looking at (\ref{mde3}), for the co-ordinate transformation $u = r^{-1}$, we see that (\ref{orgnl}) is 
dual to 
$$\left( \frac{d u}{d \psi} \right)^2 + u^2 = A u^n + B.$$
The dual versions of the equations above are identical in form only for $n=4$, which 
according to (\ref{pots}), for $f(r) = g(r)$ and $G(r) \propto r^{2-n}$ means that $V(r) \propto r^{-4}$. This 
shows that null geodesics for Schwarzschild-Tangherlini metrics for $n = 4$ exhibit self-
dual orbits under co-ordinate inversion $(r, \phi) \leftrightarrow (u = r^{-1}, \psi = - \phi)$.

\item For $p = -\frac12$, we get what will be the focus of our discussion.
$$V (|z|) \propto |z|^{-3}, \qquad \qquad \widetilde{V} (|w|) = |w|^{-6}.$$ 
This conformal map is essentially the combination of the above two conformal maps as 
we shall describe below. If we write the conformal map $w = z^{-\frac12}$, it essentially means $z = w^{-2}$.
Here, we shall compare the equations for conformal transformations $(p = 2, n = 6)$ and 
$(p=-1, n = 3)$. Let us define a co-ordinate $u = r^{-1}$, such that we have the conformal 
co-ordinate map 
\begin{equation}
\label{map1} ( y, \varphi ) \longrightarrow \left( \dfrac1{u^2}, - 2 \psi \right).
\end{equation}
to (\ref{mde3}) for the case of $n = 3$. This will give us:
$$\left( \frac{d y}{d \varphi} \right)^2 + y^2 = 2 M_n y^3 + \frac1{b^2}.$$
which under the conformal map (\ref{map1}) transforms into 
$$\left( \frac{d u}{d \psi} \right)^2 + u^2 = 2 M_n + \frac{u^6}{b^2} = 2 \widetilde{M}_n u^6 + \frac1{\widetilde{b}^2}.$$
Showing that the cases $n=3$ and $n=6$ are dual to each other, provided we redefine the 
coefficients as $2 \widetilde{M}_n = \frac1{b^2}, \widetilde{b}^2 = \frac1{2 M_n}$.
\end{enumerate}
Thus, null-geodesics derived from the Schwarzschild-Tangherlini metric under Bohlin transformation 
for $n=6$, and under co-ordinate inversion for $n=3$ produce dual equations.

\section{Conclusion and Discussion}

We managed to describe spacetimes with gravitational fields as optically refractive media, 
and elaborately reformulated mechanics in the classical limit in an optical-mechanical form. 
However, we have shown that the Hamilton-Jacobi equation for optical mechanical formulation 
was not comparable to the Eikonal equation as claimed in \cite{br, clanc, eni1, eni2}. 

We have also shown that isotropic spacetimes where the spatial part of the metric is not 
flat exhibit mechanics with a drag force quadratically dependent on velocity as seen in its 
equations of motion. However, when considering the Gorringe-Leach equations, we are dealing 
with a drag force with a different form of quadratic dependence on velocity. This results in a 
corresponding spacetime that has the same refractive index as without the drag.

It was shown that null-geodesics can be recast as central force mechanical systems, which allows us 
to deduce their trajectories deduced via existing solutions to Binet's equation in dynamics. For suitable 
approximations, mainly involving large radii, we also showed that the null-geodesic solutions can be 
deduced in terms of incomplete Beta functions, with computation of the deflection angle involving the 
familiar beta functions. Furthermore, for two choices of metric co-efficient functions, we can get systems 
with drag like Helmholtz \cite{almsan} and Helmholtz-Duffing \cite{asyyky} oscillators, and Binet's equations to Tomimatsu-
Sato and Kerr spacetimes. 

We paid special attention to the study of the Tomimatsu-Sato metric where we examined the limits, 
and solved for the null geodesics in the equatorial plane in those cases. The weak field limit, disc model, 
and the limiting case of $q = 1$ were analyzed, and the null-geodesic solutions were deduced in each of the 
limiting cases, including only one setting of the $q = 1$ limit.

Finally, we explored dual systems that preserve the classic Jacobi metric under conformal transformation, 
one of which was the Bertrand system pair.  

\section*{Acknowledgement}
We are grateful to Professor Donato Bini for his suggestion  to work on 
Tomimatsu-Sato metric and numerous discussions.
We also thank Prof. Gary Gibbons, Prof. Marco Cariglia, and Dr. Francesco Messina for their correspondence 
and advice.  PG is grateful to IHES, 
and IFSC, Sao Paulo where part of the work was done.
 The research of PG was partly supported by FAPESP through 
Instituto de Fisica de S\~ao Carlos, Universidade de Sao Paulo with grant number 2016/06560-6.


\begin{thebibliography}{999}

\bibitem{ghww} G.W. Gibbons, C.A.R. Herdeiro, C.M. Warnick, and M.C. Werner, {\it Stationary Metrics and Optical Zermelo-Randers-Finsler Geometry}, \href{http://journals.aps.org/prd/abstract/10.1103/PhysRevD.79.044022}{Phys. Rev. D. {\bf 4} (2009) 79}, arXiv: \href{http://arxiv.org/abs/0811.2877}{0811.2877}.

\bibitem{casey2} S. Casey, {\it Kastor-Traschen Black Holes, Null Geodesics and Conformal Circles}, \href{http://iopscience.iop.org/article/10.1088/0264-9381/29/13/135006/meta}{Class. Quantum Grav. {\bf 29} (2012) 135006}, arXiv: \href{https://arxiv.org/abs/1110.2091}{1110.2091}.

\bibitem{maup} P.L. Maupertuis, (1744),  \href{https://fr.wikisource.org/wiki/Accord_de_diff%C3%A9rentes_loix_de_la_nature_qui_avoient_jusqu%E2%80%99ici_paru_incompatibles}{\it Accord de diff\'erentes loix de la nature qui avoient jusqu'ici paru incompatibles}, Acad\'emie Internationale d'Histoire des Sciences, Paris.

\bibitem{gwg} G.W. Gibbons, {\it The Jacobi-metric for timelike geodesics in static spacetimes}, \href{http://iopscience.iop.org/article/10.1088/0264-9381/33/2/025004/meta}{Class. Quantum Grav. {\bf 33} (2015) 025004}, arXiv: \href{http://arxiv.org/abs/1508.06755}{1508.06755}.

\bibitem{cgg} S. Chanda, G.W. Gibbons and P. Guha, {\em Jacobi-Maupertuis-Eisenhart metric and geodesic flows}, \href{http://aip.scitation.org/doi/full/10.1063/1.4978333}{J. Math. Phys. {\bf 58} (2017) 032503}, arXiv: \href{https://arxiv.org/abs/1612.00375}{1612.00375}.

\bibitem{cgg1} S. Chanda, G.W. Gibbons and P. Guha, {\em Jacobi-Maupertius metric and Kepler equation},
\href{http://www.worldscientific.com/doi/abs/10.1142/S0219887817300021?journalCode=ijgmmp}{Int. J. Geom. Methods in Mod. Phys. {\bf 14} (Iss. 07) (2017) 1730002}, arXiv: \href{https://arxiv.org/abs/1612.07395v1}{1612.07395 [math-ph]}.

\bibitem{gwg1}  G.W. Gibbons and M. Vyska, {\em The applications of Weierstrass elliptic functions to Schwarzschild
null geodesics}, \href{http://iopscience.iop.org/article/10.1088/0264-9381/29/6/065016/meta}{Class. Quantum Grav. {\bf 29} (2012) 065016 (19 pp)}, arXiv: \href{https://arxiv.org/abs/1110.6508}{1110.6508}.


\bibitem{dual} Y. Grandati, A. B\'erard and H. Mohrbach, {\it Bohlin-Arnold-Vassiliev's Duality and Conserved Quantities}, arXiv: \href{https://arxiv.org/abs/0803.2610}{0803.2610v2 [math-ph]}, 1-8.

\bibitem{mls} M.L. Saggio, {\em Bohlin transformation: the hidden symmetry that connects Hooke to Newton}, \href{http://iopscience.iop.org/0143-0807/34/1/129}{Eur. J. Phys. {\bf 34} (2013) 129}.

\bibitem{casey1} S. Casey, {\it Optical 2-metrics of Schwarzschild-Tangherlini spacetimes and the Bohlin-Arnold duality}, \href{http://iopscience.iop.org/article/10.1088/0264-9381/29/23/237001/meta}{Class. Quantum Grav. {\bf 29} (2012) 237001}, arXiv: \href{https://arxiv.org/abs/1208.0168v1}{1208.0168 [gr-qc]}.

\bibitem{cgly} S. Chen, G.W. Gibbons, Y. Li, and Y. Yang, {\em Friedmann's equations in all dimensions and Chebyshev's theorem}, \href{http://iopscience.iop.org/article/10.1088/1475-7516/2014/12/035/meta}{J. Cosmol. Astropart. Phys. 2014 {\bf 12} (2014) 035}, arXiv: \href{https://arxiv.org/abs/1409.3352v2}{1409.3352 [astro-ph]}.

\bibitem{mz} E.A. Marchisotto, and G.A. Zakeri, {\em An invitation to integration in finite terms}, \href{http://www.jstor.org/stable/2687614?origin=JSTOR-pdf}{1994 College Math. J. {\bf 25} (1994) 295-30}.

\bibitem{plmath} \href{http://planetmath.org/integrationofdifferentialbinomial}{http://planetmath.org/integrationofdifferentialbinomial}

\bibitem{KR} S. Khorasani and B. Rashidian, {\em Optical anisotropy of schwarzschild metric within 
equivalent medium framework}, \href{http://www.sciencedirect.com/science/article/pii/S0030401809012711}{Opt. Commun. {\bf 283} (2010) 1222-1228}, arXiv: \href{https://arxiv.org/abs/0902.0728}{0902.0728}.

\bibitem{tomsato} A. Tomimatsu, and H. Sato, {\em New exact solution for the gravitational field of a spinning mass}, \href{https://journals.aps.org/prl/abstract/10.1103/PhysRevLett.29.1344}{Phys. Rev. Lett. {\bf 29} (1972) 1344}.

\bibitem{tomsatop} A. Tomimatsu, and H. Sato, {\em New series of exact solutions for gravitational fields of spinning masses}, \href{https://academic.oup.com/ptp/article/50/1/95/1898929}{Progr. Theor. Phys. {\bf 50} (1973) 95-110}.

\bibitem{bosewang} S.K. Bose, and M. Y. Wang, {\em Geodesic Motions in the Tomimatsu-Sato Metric}, \href{https://journals.aps.org/prd/abstract/10.1103/PhysRevD.8.361}{Phys. Rev. D {\bf 8} (1973) 361}.

\bibitem{Chebyshev} P.L.Chebyshev, {\em Sur l'int ́egration des diff\ ́\'erentielles irrationnelles}, 
J. Math. Pures Appl., (1853),
v.18, pp. 87-111; Oeuvres vol. 1, pp. 147-168.

\bibitem{kinkel} W. Kinnersley, and E. F. Kelley, {\em Limits of the Tomimatsu-Sato gravitational field}, \href{aip.scitation.org/doi/abs/10.1063/1.1666592}{J. Math. Phys. {\bf 15} (1974) 2121-2126}.

\bibitem{glass} E.N. Glass, {\em Structure of the tomimatsu-sato gravitational field}, \href{https://journals.aps.org/prd/abstract/10.1103/PhysRevD.7.3127}{Phys. Rev. D {\bf 7} (1973) 3127}.

\bibitem{br} A.M. Bloch and A.G. Rojo, {\em Optical mechanical analogy and nonlinear nonholonomic constraints}, \href{http://journals.aps.org/pre/abstract/10.1103/PhysRevE.93.023005}{Phys. Rev. E Stat Nonlin Soft Matter Phys}.

\bibitem{fnb} I. Fern\'andez-N\'u\~nez and O. Bulashenko, {\em Anisotropic metamaterial as an analogue of a black hole}, \href{http://www.sciencedirect.com/science/article/pii/S0375960115009159}{Phys. Lett. A {\bf 380} (2016) 1-8}, arXiv: \href{https://arxiv.org/abs/1507.08152}{1507.08152}.

\bibitem{cs} W. Cai and V. Shalaev, {\em Optical Metamaterials: Fundamentals and Applications}, \href{https://link.springer.com/book/10.1007/978-1-4419-1151-3}{(Springer, 2010)}.

%\bibitem{csl} T. J. Cui, D. R. Smith, and R. Liu, {\em Metamaterials: Theory, Design, and Applications} \href{https://link.springer.com/book/10.1007/978-1-4419-0573-4}{(Springer, New York, 2010)}.

\bibitem{csl} D. R. Smith, J.B. Pendry and M.C.K. Wiltshire, {\em Metamaterials and negative refractive 
index}, Science {\bf 305} (2004) - science.sciencemag.org.

\bibitem{bgj} D. Bini, A. Geralico and R.T. Jantzen. {\em Frenet-Serret formalism for null world lines}, \href{http://iopscience.iop.org/article/10.1088/0264-9381/23/11/018/meta}{Class. Quantum Grav. {\bf 23} (2006) 3963}.

\bibitem{cg} S. Chanda, and P. Guha, {\em Geometrical Formulation of Relativistic Mechanics}, Int. J. Geom. Methods in Mod. Phys. {\bf 15} (2018) 1850062, arXiv: \href{https://arxiv.org/abs/1706.01921}{1706.1921 [hep-th]}.

\bibitem{cgcg} S. Chanda, A. Ghose-Choudhury, and P. Guha, {\em Jacobi-Maupertuis metric of Lienard type equations and Jacobi Last Multiplier}, arXiv:\href{https://arxiv.org/abs/1706.02219}{1706.02219}.

\bibitem{alsing} P. M. Alsing, {\em The optical-mechanical analogy for stationary metrics in general relativity}, \href{http://aapt.scitation.org/doi/abs/10.1119/1.18957}{Am. J. Phys. {\bf 66} (1998) 779}.

\bibitem{clanc} C. Lanczos, {\em The variational principles of mechanics}, Courier Corporation, 2012.

\bibitem{gbm} Y. Grandati, A. B\'erard, and H. Mohrbach, {\em Duality properties of Gorringe-Leach equations}, \href{https://link.springer.com/article/10.1007/s10569-008-9174-1}{Celest. Mech. Dyn. Astron. {\bf 103}, no. 2 (2009) 133-141}.

\bibitem{cdgh} M. Cariglia, C. Duval, G.W. Gibbons and P.A. Horv\'athy, {\em Eisenhart lifts and symmetries of time-dependent systems}, \href{http://www.sciencedirect.com/science/article/pii/S0003491616301348}{Ann. Phys. {\bf 373} (2016) 631-654}, arXiv: \href{https://arxiv.org/abs/1605.01932}{1605.01932}.

\bibitem{almsan} J.A. Almendral and M.A.F. Sanju\'an, {\em Integrability and symmetries for the Helmholtz oscillator with friction}, \href{http://www.s512731021.mialojamiento.es/pdfs/almendral2003b.pdf}{J. Phys. A: Math. Gen. {\bf 36} (2003) 695-710}.

\bibitem{asyyky} H. Askari, Z. Saadatnia, D. Younesian, A. Yildirim, M. Kalami-Yazdi, {\em Approximate periodic solutions for the Helmholtz-Duffing equation}, \href{http://www.sciencedirect.com/science/article/pii/S0898122111008108?np=y&npKey=2791864f79d1d9af733ffcfa573435f25df87f0e285af8ec56ab2eadae534001}{Computers \& Mathematics with Applications {\bf 62} (2011) 3894-3901}.

\bibitem{cvet} L. Cveticanin, {\em Vibrations of the nonlinear oscillator with quadratic nonlinearity}, \href{http://www.sciencedirect.com/science/article/pii/S0378437104005278}{Physica A: Statistical Mechanics and its Applications {\bf 341} (2004) 123-135}.

\bibitem{cvet1}  I. Kovacic, L. Cveticanin, M, Zukovic and Z. Rakaric, {\em Jacobi elliptic functions: A review of nonlinear oscillatory application problems}, \href{http://www.sciencedirect.com/science/article/pii/S0022460X16302279}{J. Sound Vibration {\bf 380} (2016) 1-36}.

\bibitem{tomsato1} A. Tomimatsu, and H. Sato, {\em Event horizon of the Tomimatsu-Sato metrics}, Lettere Al Nuovo Cimento (1971–1985) {\bf 8} (1973) 740-742.

\bibitem{tomsato2} M. Yamazaki, {\em On the Kerr and the Tomimatsu-Sato Spinning Mass Solutions}, Progr. Theor. Phys. {\bf 57} (1977) 1951-1957.

\bibitem{hikod} W. Hikida, and H. Kodama, {\em An Investigation of the Tomimatsu-Sato Spacetime}, arXiv: \href{https://arxiv.org/abs/gr-qc/0303094v1}{gr-qc/0303094}.

\bibitem{mcar} M. Cariglia, {\it Null lifts and projective dynamics}, \href{http://www.sciencedirect.com/science/article/pii/S0003491615003371}{Ann. Phys. {\bf 362} (2015) 642-658}, arXiv: \href{https://arxiv.org/abs/1506.00714}{1506.00714}.

\bibitem{bohlin} M.K. Bohlin, {\it Note sur le probl\'eme des deux corps et sur une int\'egration nouvelle dans le probl\'eme des trois corps}, \href{http://adsabs.harvard.edu/full/1911BuAsI..28..113B}{Bulletin Astronomique, Serie I {\bf 28} (1911) 113-119}.

\bibitem{arnold} V.I. Arnold, (1990), Huygens and Barrow, Newton and Hooke, Basel: Birkhauser.

\bibitem{eni1} J. Evans, K.K. Nandi and A. Islam, {\em The optical-mechanical analogy in general relativity: New methods for the paths of light and of the planets}, \href{http://aapt.scitation.org/doi/abs/10.1119/1.18366}{Am. J. Phys. {\bf 64} (1996) 1404}.

\bibitem{eni2} J. Evans, K.K. Nandi and A. Islam, {\em The optical-mechanical analogy in general relativity: exact 
Newtonian forms for the equations of motion of particles and photons}, \href{https://link.springer.com/article/10.1007/BF02105085}{Gen. Rel. Grav. {\bf 28} (1996) 413-439}.


\end{thebibliography}
\end{document}